%% file: main.tex
\documentclass{article}

\usepackage{PRIMEarxiv}

\usepackage[utf8]{inputenc} 
\usepackage[T1]{fontenc}    
\usepackage{hyperref}       
\usepackage{url}            
\usepackage{booktabs}       
\usepackage{amsfonts}       
\usepackage{nicefrac}       
\usepackage{microtype}      
\usepackage{fancyhdr}       
\usepackage{graphicx}       
\graphicspath{{figs/}}     

\usepackage{svg}
\usepackage{graphicx}
\usepackage{bm}
\usepackage{newtxtext}
\usepackage{newtxmath}
\usepackage{natbib}
\usepackage{multirow}
\usepackage{hyperref}
\usepackage{cleveref}
\hypersetup{
    colorlinks = true,
    urlcolor   = blue,
    citecolor  = black,
}

{\left\lbrace\begin{aligned}{{}{}}}%
{\end{aligned}\right.}
\newcommand{\RomanNumeralCaps}[1]
\linenumbers
\usepackage{comment}
\usepackage{mathtools}

\usepackage{authblk}

\title{Quasi-linear homogenization for large-inertia laminar transport across permeable membranes}

\author[1]{K. Wittkowski}
\author[2]{A. Ponte}
\author[3]{P.G. Ledda}
\author[1]{G.A. Zampogna*}

\affil[1] {LFMI, \'Ecole Polytechnique F\'ed\'erale de Lausanne, CH-1015 Lausanne, Switzerland}
\affil[2]{DICCA, Università degli Studi di Genova, Via Montallegro 1, 16145 Genova, Italy}
\affil[3]{DICAAR, Università degli Studi di Cagliari, Via Marengo 2, 09123 Cagliari, Italy}

\begin{document}
\maketitle
\begingroup\def\thefootnote{*}\footnotetext{Corresponding author e-mail: giuseppe.zampogna@epfl.ch}\endgroup

\begin{abstract}
\textcolor{black}{Porous membranes are thin solid structures that allow the flow to pass through their tiny openings, called pores. Flow inertia may play a significant role in several filtration flows of natural and engineering interest. Here, we develop a predictive macroscopic model to describe solvent and solute flows past thin membranes for non-negligible inertia. We leverage homogenization theory to link the solvent velocity and solute concentration to the jumps of solvent stress and solute flux across the membrane. Within this framework, the membrane acts as a boundary separating two distinct fluid regions. These jump conditions rely on several coefficients, stemming from closure problems at the microscopic pore scale. Two approximations for the advective terms of Navier-Stokes and advection-diffusion equations are introduced to include inertia in the microscopic problem. The approximate inertial terms couple the micro- and macroscopic fields. Here, this coupling is solved numerically using an iterative fixed-point procedure. We compare the resulting models against full-scale simulations, with a good agreement both in terms of averaged values across the membrane and far-field values. Eventually, we develop a strategy based on unsupervised machine learning to improve the computational efficiency of the iterative procedure. The extension of homogenization toward weak-inertia flow configurations as well as the performed data-driven approximation may find application in preliminary analyses as well as optimization procedures toward the design of filtration systems, where inertia effects can be instrumental in broadening the spectrum of permeability and selectivity properties of these filters.}
\end{abstract}


\input{1-introduction}
\input{2-model}

\input{3A-micro_results}

\input{3B-macro_results}

\input{3C-computat_efficiency}
\input{4-conclusion}
\input{5-acknowledgements}
\input{6-code}

\bibliographystyle{unsrt}  
\bibliography{ref}  

\input{Appendix}

\end{document}

%% file: 1-introduction.tex
\section{Introduction}\label{sec:intro}
Flows across thin permeable structures are of considerable interest in a wide range of natural and engineering situations. In nature, nocturnal birds of prey like owls owe their silent flight to the porous texture of the feathers covering their wings \citep{peake2020}.
The pappus, a parachute-like porous structure, can stabilise and decelerate the glide of dandelion seeds \citep{cummins2018separated} thanks to a separated vortex ring \citep{ledda2019}. 
Starting from the previous example, a relevant application where inertial flows across membranes take place are parachutes. The flow around these devices has been extensively studied both experimentally and numerically \citep{payne_1978,greathouse2015,bergeron2022}, finding relevant dependencies between the parachute effectiveness and the porous microstructure.
Deep-sea porous sponges like \textit{Euplectella Aspergillum} have recently attracted attention not only thanks to their remarkable structural resistance but also for their ability to optimise their internal re-circulation patterns, which in turn are beneficial to the organisms living inside this glass-sponge \citep{Falcucci2021}. In the latter case, the filtration properties of these sponges are crucial in defining these recirculating patterns as well as the transport of nutrients within their structures.

Microfluidics and filtration systems are widely employed in separation and filtration processes at the cell scale \citep{catarino2019blood} and beyond.
The flow is usually laminar but, in certain applications, advection may play an important role in the transport of solute \citep{Tripathi_2015}. Advection effects are quantified through two non-dimensional numbers: the Reynolds number, the ratio between inertial and viscous scales for the fluid flow, and the Péclet number, the ratio between the advective and diffusive scales for the transport of solute in a solvent. In membrane filters for the collection of particles, the Reynolds number can reach values up to $20$ \citep{yang1999}. In hydrogen fuel cells, the Péclet number inside a proton-exchange membrane can be of order $10$ \citep{Suresh2016}. Microfluidic mixing processes can be enhanced by adding microstructured patterns within the micro-channels \citep{stroock2002}. However, filtration processes do not only involve lab-scale systems. As a matter of fact, nets and harps are attracting interest in harvesting water from fog in arid environments \citep{park2013optimal,labbe2019capturing, MONCUQUET2022106312}.

Accurately modelling the transport phenomena across porous membranes in the presence of inertial effects thus affects a variety of applications ranging across several length and time scales. Numerical studies concerning the interaction between fluid flows and permeable structures belong to two approaches: full-scale solutions and averaged models. Direct full-scale solutions such as the one reported in \cite{icardi2014}, although very accurate, require a non-trivial computational effort from the geometry and mesh generation to the actual numerical solution. This approach rapidly becomes prohibitive for flows at industrial scales or in the case of biological systems with extreme scale separation, e.g. cell membranes, where pores are nanometric and membranes are micrometric \citep{verkman2000}. In addition, full-scale solutions offer sometimes little insight into the general physics governing the processes and they are not scalable in the case of parametric studies or optimization routines. Conversely, simplified models are preferred in some contexts. 
Some authors studied analytically the specific case of flow normal to the membrane \citep{conca2,bourgeat3}, while others developed theories for flow through infinitesimally thin porous membranes and deduced some range of applicability of the Stokes' approximation \citep{Tio1994} or considered simplified pore geometries and arrangements \citep{jensen2014flow}. A specific class of simplified flow descriptions consists of averaged models. 
Early models describe the fluid flow and solute transport across a bulk porous medium with a solvent flow description analogous to Darcy law (\citealp{darcy}, see \cite{dagan1987theory} for a review). Despite being computationally less expensive, these models converge to the actual fluid flow field only in an average sense and rely on empirical coefficients, like the permeability, difficult to quantify from a theoretical point of view, thus limiting the model's predictive power. Multi-scale techniques such as the volume-averaged method \citep{Whitaker1996} and homogenization \citep{hornung} enable a more accurate and predictive description of such flows. In homogenization, the medium properties are the spatially-averaged solution of closure problems at the pore-scale level. Homogenization was shown to be a valid methodology for predicting inertia-less fluid flows. \cite{zampogna2020effective} accurately predicted the permeability tensor of thin porous membranes through a formal approach, including the transport of diluted species \citep{zampogna2022transport}. \cite{ledda2021homogenization} showed that this methodology can be used not only for flow analysis but also for membrane design and optimization. Homogenization is thus established as a formal approach for studying and designing porous structures when the flow at the pore scale has negligible inertia. However, at the current state of the art, homogenization cannot handle pore-scale inertial flows since it requires the linearity of the governing pore-scale equations.  
To overcome this limitation, \cite{zampogna_canopy_instability} and \cite{luminari2018} proposed an Oseen-like momentum equation to close the pore-scale problem in the case of flows through bulk porous media. 

Actual applications in thin membrane flows can significantly benefit from extending beyond the inertia-less regime the modelling and optimization strategies associated with homogenization to better upscale transport and filtration phenomena occurring at the pore scale. In the present work, we generalize the framework proposed in \cite{zampogna2022transport} for inertial pore-scale flows. In \S \ref{sec:hommodel} we present the mathematical derivation of the homogenization procedure. Section \ref{sec:microfields} presents the solution to the problems we solve at the microscopic level to obtain the effective permeability and diffusivity coefficients, while in \S \ref{sec:fullscalemacro} we use these solutions to predict the mean flow behaviour past permeable membranes. We compare our homogeneous model with simulations solved at all scales. In \S \ref{sec:computeffic} we improve the computational efficiency of our methodology by introducing a machine-learning algorithm to minimize the number of microscopic problems to be solved. In \S \ref{sec:conclu} we discuss our results and future perspectives. 

%% file: 2-model.tex
\section{Homogenized model and quasi-linear inertial flow extension}\label{sec:hommodel}
\begin{figure}
\centering
    \includegraphics[width=\textwidth]{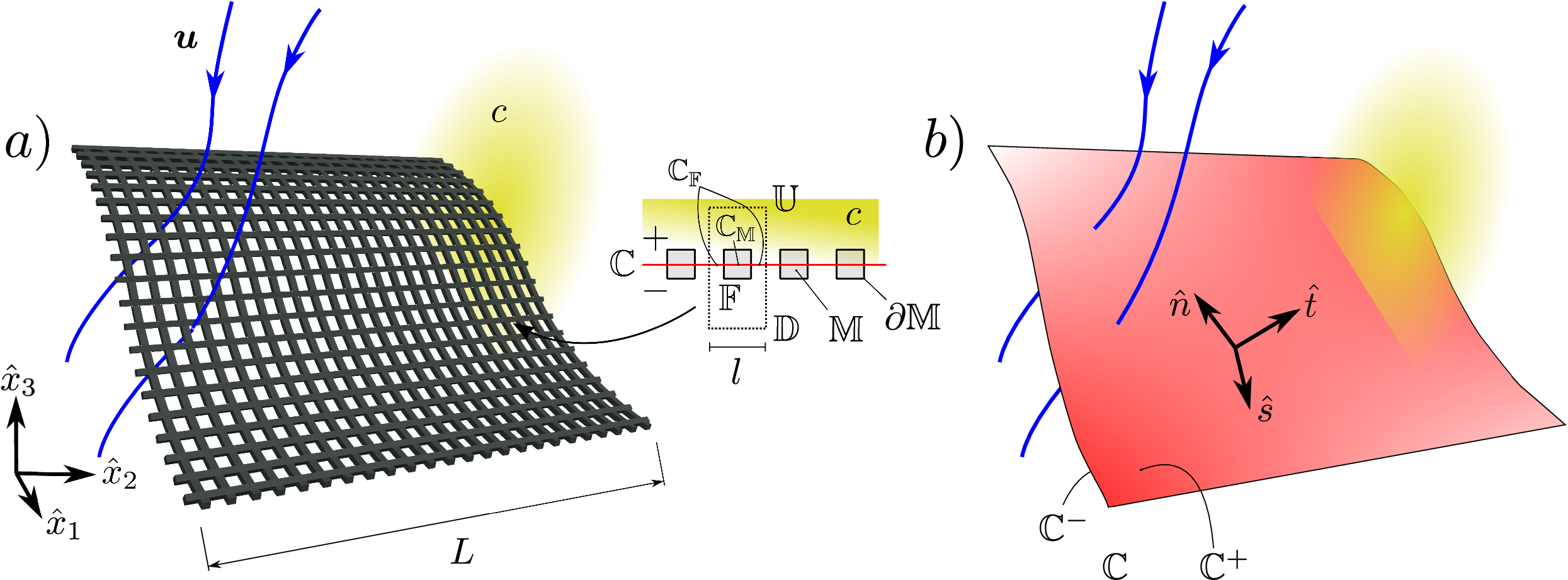}
    \caption{Porous thin membrane. Panel $a$ shows the membrane as it appears in the physical world. The fluid flow (blue streamlines) crosses the membrane via its pores and the concentration field (yellow contour) interacts with the membrane surface at the pore scale. Panel $b$ depicts a purely macroscopic domain, where the membrane is substituted by its mean surface $\mathbb{C}$ and where the details of the pores have been coarse-grained: $\mathbb{C}$ is a fictitious surface, separating two domains, where equivalent boundary conditions are imposed to reproduce the average effect of the membrane on the fluid flow and concentration fields. $\mathbb{C_F}$ and $\mathbb{C_M}$ are the portions of $\mathbb{C}$ occupied by the fluid and by the solid, respectively}
    \label{fig:generaldomain}
\end{figure}

\begin{figure}
\centering
    \includegraphics[width=\textwidth]{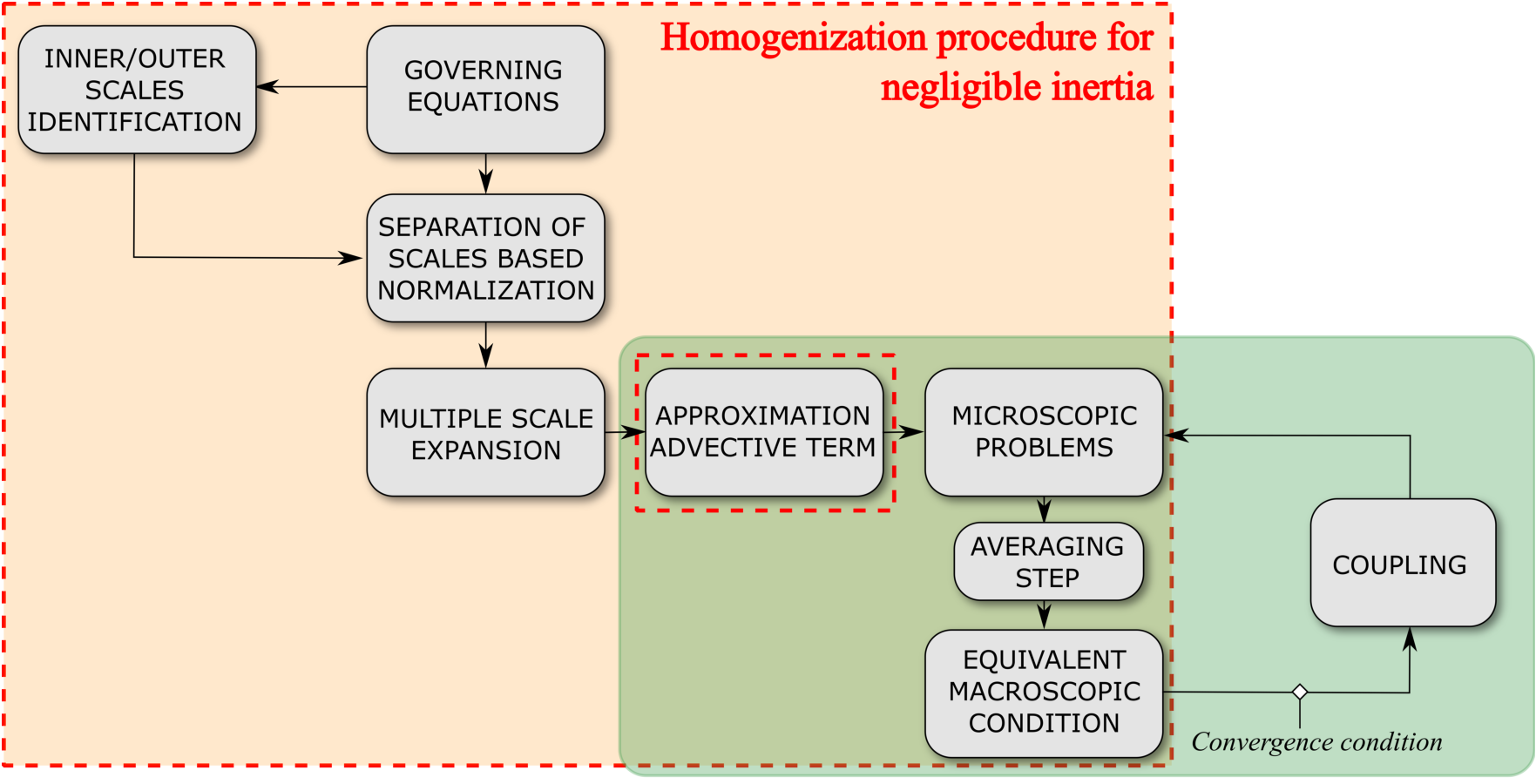}
    \caption{Diagram of the procedure used to deduce the macroscopic model. The green box highlights the iterative part of the procedure.}
    \label{flow:procedure}
\end{figure}

We consider the incompressible flow of a Newtonian fluid (so-called solvent) of density $\rho$ and viscosity $\mu$ travelling across a thin microstructured porous membrane and transporting a diluted solute of diffusivity $D$. We introduce the solute concentration $\hat{c}$ and the solvent velocity and pressure fields $\hat{u}_i,\hat{p}$. The fluid domain and porous structure are depicted in figure \ref{fig:generaldomain}.  
Denoting as $l$ and $L$ the pore (micro-scale) and the membrane (macro-scale) length scales, respectively, we define the separation of scales parameter as $\epsilon=l/L$. The full-scale problem is governed by the Navier-Stokes equations for the solvent velocity and pressure, as well as by the advection-diffusion equation for the solute concentration, 
\begin{equation}
    \begin{cases}                               
    \rho(\hat{\partial}_{\mathsf{t}}\hat{u}_i+\hat{u}_j\hat{\partial}_j \hat{u}_i)=-\hat{\partial}_i \hat{p}+\mu \hat{\partial}_{jj}^2 \hat{u}_i,
    \\
    \hat{\partial}_i \hat{u}_i=0,
    \\
    \hat{\partial}_{\mathsf{t}}\hat{c}=-\hat{u}_i \hat{\partial}_i \hat{c}+D\hat{\partial}^2_{ll} \hat{c},
    \label{eq:nsedim}\end{cases}
\end{equation}
where the Einstein index notation is adopted. As shown in the following sections, homogenization relies on the following steps (figure \ref{flow:procedure}): $(i)$ define the \textit{inner} equations and normalization, which apply at the pore scale, $(ii)$ define the \textit{outer} equations and normalizations, which apply far from the membrane, $(iii)$ match the inner and outer domains, $(iv)$ solve the inner (microscopic) problem, $(v)$ average the inner solution and deduce the macroscopic condition.
\subsection{The inner problem}\label{sub:2innerproblem}
We refer to the domain $\mathbb{F}$ in figure \ref{fig:generaldomain}a as the \textit{microscopic} domain, in opposition to the outer, \textit{macroscopic} domain, formed by the fluid region far from the membrane (figure \ref{fig:generaldomain}b). In the inner domain, we assume that the flow is dominated by viscous dissipation, which leads to the following scaling
\begin{equation}
     \hat{p}=\Delta \mathcal{P} p=\frac{\mu \mathcal{U}}{l}p \qquad \hat{u}=\mathcal{U}u \qquad \hat{\mathsf{t}}=T\mathsf{t}=\frac{l}{\mathcal{U}}\mathsf{t} \qquad \hat{c}=Cc,
     \label{eq:scalemicro}
\end{equation}
where $\Delta \mathcal{P}, \mathcal{U}, T$ and $C$ are the scales of pressure difference, velocity, time and concentration at the pore level, respectively. The equations governing the physics within the microscopic elementary cell, $\mathbb{F}$, are
\begin{equation}
    \begin{cases}                               
    Re_l(\partial_{\mathsf{t}}u_i+ u_j\partial_j u_i)=-\partial_i p+ \partial_{jj}^2 u_i,
    \\
    \partial_i u_i=0,
    \\
    Pe_l\partial_{\mathsf{t}} c=-Pe_l u_i \partial_i c+\partial^2_{ll} c,
    \label{eq:nseadeinner}\end{cases}
\end{equation}
where $Pe_l=\mathcal{U}l/D$ and $Re_l=\mathcal{U}l/\nu$ are the Péclet and Reynolds numbers referred to the microscopic length $l$, respectively. The flow is assumed to be periodic along the tangential-to-the-membrane directions. No-slip ($u_i=0$) and chemostat-like ($c=0$) boundary conditions are imposed on the fluid-solid interface $\partial \mathbb{M}$. These Dirichlet boundary conditions are contained in the more general set of Robin boundary conditions presented in \cite{zampogna2022transport}.  
\subsection{The outer problem}\label{sub:2outerproblem}
In the macroscopic domain, we assume that the flow is inertia-dominated, i.e. the following non-dimensionalization is employed to scale the equations:
\begin{equation}
     \hat{p}=\Delta \mathcal{P}^{\mathbb{O}} p^{\mathbb{O}}=\rho {\mathcal{U}^{\mathbb{O}}}^2 p^{\mathbb{O}} \qquad \hat{u}=\mathcal{U}^{\mathbb{O}} u^{\mathbb{O}} \qquad \hat{\mathsf{t}}=T^{\mathbb{O}} \mathsf{t}^{\mathbb{O}}=\frac{L}{\mathcal{U}^{\mathbb{O}}}\mathsf{t}^{\mathbb{O}} \qquad \hat{c}=C^{\mathbb{O}}c^{\mathbb{O}}.
     \label{eq:scalemacro}
\end{equation}
The governing equations of the outer problem are
\begin{equation}
    \begin{cases}                               
   \partial_{\mathsf{t}}u_i^{\mathbb{O}}+u_j^{\mathbb{O}}\partial_j u_i^{\mathbb{O}}=-\partial_i p^{\mathbb{O}}+ \frac{1}{Re_L}\partial_{jj}^2 u^{\mathbb{O}}_i,
    \\
    \partial_i u^{\mathbb{O}}_i=0,
    \\
    Pe_L\partial_{\mathsf{t}} c^{\mathbb{O}}=-Pe_L u_i^{\mathbb{O}} \partial_i c^{\mathbb{O}}+\partial^2_{ll} c^{\mathbb{O}},
    \label{eq:nseadeouter}\end{cases}
\end{equation}
where $Pe_L=\mathcal{U}^{\mathbb{O}}L/D$ and $Re_L=\mathcal{U}^{\mathbb{O}}L/\nu$.
\subsection{Matching the inner and outer domains}\label{sub:2matching}
The ratio of microscopic and macroscopic time scales,
\begin{equation}
    \frac{T}{T^{\mathbb{O}}}=\frac{l}{\mathcal{U}}\frac{\mathcal{U}^{\mathbb{O}}}{L}=\epsilon \frac{\mathcal{U}^{\mathbb{O}}}{\mathcal{U}},
\end{equation}
suggests that the variation of the micro-scale occurs in a much smaller time compared to the characteristic time variations at the macro-scale, thus the pore-scale problem can be considered steady if no unsteadiness is triggered at the level of the micro-scale,
\begin{equation}
    \begin{cases}                               
    Re_l u_j\partial_j u_i=-\partial_i p+ \partial_{jj}^2 u_i,
    \\
    \partial_i u_i=0,
    \\
    -Pe_l u_i \partial_i c+\partial^2_{ll} c=0.
    \label{eq:nseadeinnerEQ}\end{cases}
\end{equation}
 On the upward $\mathbb{U}$ and downward $\mathbb{D}$ sides of $\mathbb{F}$, the outer and inner fluid flow and concentration fields match, i.e. 
\begin{equation}
    \Sigma_{jk}n_k=\epsilon Re_L \frac{U^{\mathbb{O}}}{U}\Sigma_{jk}^{\mathbb{O}}n_k \qquad u=\frac{U^{\mathbb{O}}}{U}u^{\mathbb{O}} \qquad F_j n_j=\frac{C^{\mathbb{O}}}{\epsilon C}F_j^{\mathbb{O}}n_j \qquad c=\frac{C^{\mathbb{O}}}{C}c^{\mathbb{O}},
\label{eq:bcdirect}\end{equation}
where $\Sigma_{jk}=-p\delta_{jk}+(\partial_j u_k+\partial_k u_j)$, $\Sigma_{jk}^{\mathbb{O}}=-p^{\mathbb{O}}\delta_{jk}+\frac{1}{Re_L}(\partial_j u^{\mathbb{O}}_k+\partial_k u^{\mathbb{O}}_j)$ ,  $F_j=Pe_l u_i c - \partial_i c$  and $F_j^{\mathbb{O}}=Pe_L u_i^{\mathbb{O}} c^{\mathbb{O}} - \partial_i c^{\mathbb{O}}$ are the fluid stresses and solute fluxes in the inner and outer domains, respectively. 
\subsection{Solving the inner problem}\label{sub:2solvinner}
To apply homogenization to problem \eqref{eq:nseadeinnerEQ}, the Stokes approximation assumes $Re_l\sim Pe_l\sim \epsilon$ \citep{zampogna2022transport}. In this paper, we introduce finite Péclet and Reynolds numbers at the pore scale. As a consequence, problem \eqref{eq:nseadeinnerEQ} is a set of non-linear PDEs. Exploiting the separation of scales, we perform the following asymptotic expansion:
\begin{equation}
    x_i=x_i+\epsilon X_i \quad \partial_i =\partial_i+\epsilon \partial_I \quad (u_i,p,c)=(u_i^{(0)},p^{(0)},c^{(0)})+\epsilon (u_i^{(1)},p^{(1)},c^{(1)})+\mathcal{O}(\epsilon^2)
\label{eq:scalexpansion}\end{equation}
Substituting \eqref{eq:scalexpansion} into \eqref{eq:nseadeinnerEQ} we obtain the leading order equation
\begin{equation}
\begin{cases}
    Re_l u_j^{(0)}\partial_j u_i^{(0)}=-\partial_i p^{(0)}+ \partial_{jj}^2 u_i^{(0)},
    \\
    \partial_i u_i^{(0)}=0,
    \\
    -Pe_l u_i^{(0)} \partial_i c^{(0)}+\partial^2_{ll} c^{(0)}=0,      
\end{cases}\label{eq:micro}
\end{equation}
In order to write the solution of \eqref{eq:micro} as a linear combination of the boundary fluxes, we introduce the closure advective velocity $U_j$ such that
\begin{equation}
\begin{cases}
    Re_l U_j\partial_j u_i^{(0)}=-\partial_i p^{(0)}+ \partial_{jj}^2 u_i^{(0)},
    \\
    \partial_i u_i^{(0)}=0,
    \\
    -Pe_l U_j \partial_j c^{(0)}+\partial^2_{ll} c^{(0)}=0.      
\end{cases}\label{eq:microlin}
\end{equation}
The term $U_j$ needs to be specified to close equation \eqref{eq:microlin}. However, we refer to section \ref{sub:lineariz} for the closure of $U_j$. The solution of \eqref{eq:microlin} can be formally written as a linear combination of the boundary fluxes,
\begin{equation}
    \begin{cases}
        u_i^{(0),\mathbb{O}}=\epsilon Re_L (M_{ij}\Sigma_{jk}^{\mathbb{O,U}}n_k+N_{ij}\Sigma_{jk}^{\mathbb{O,D}}n_k) \\
        p^{(0),\mathbb{O}}=Q_{j}\Sigma_{jk}^{\mathbb{O,U}}n_k+R_{j}\Sigma_{jk}^{\mathbb{O,D}}n_k \\
        c^{(0),\mathbb{O}}=\epsilon (T F_j^{\mathbb{O,U}}n_j+S F_j^{\mathbb{O,D}}n_j),
    \end{cases}
\label{eq:model}\end{equation}
Substituting \eqref{eq:model} into \eqref{eq:microlin}, we obtain the set of solvability conditions
\begin{equation*}
    \begin{cases}
        Re_l U_m \partial_m M_{ij}=-\partial_i Q_{j} +\partial^2_{ll}M_{ij} \quad \text{in $\mathbb{F}$},\\
        \partial_i M_{ij}=0 \quad \text{in $\mathbb{F}$},\\
        \Sigma_{pq}(M_{\cdot j},Q_{j})n_q=\delta_{jp}n_q \quad \text{on $\mathbb{U}$},\\
        \Sigma_{pq}(M_{\cdot j},Q_{j})n_q=0 \quad \text{on $\mathbb{D}$},\\
        M_{ij}=0 \quad \text{on $\partial \mathbb{M}$},\\
        M_{ij},Q_{j} \quad \text{periodic along $\bm{t}$,$\bm{s}$},
    \end{cases}
    \begin{cases}
        Re_l U_m \partial_m N_{ij}=-\partial_i R_{j} +\partial^2_{ll}N_{ij} \quad \text{in $\mathbb{F}$},\\
        \partial_i N_{ij}=0 \quad \text{in $\mathbb{F}$},\\
        \Sigma_{pq}(N_{\cdot j},R_{j})n_q=\delta_{jp}n_q \quad \text{on $\mathbb{U}$},\\
        \Sigma_{pq}(N_{\cdot j},R_{j})n_q=0 \quad \text{on $\mathbb{D}$},\\
        N_{ij}=0 \quad \text{on $\partial \mathbb{M}$},\\
        N_{ij},R_{j} \quad \text{periodic along $\bm{t}$,$\bm{s}$},
    \end{cases}
\end{equation*}
\begin{equation}
    \begin{cases}
    Pe_l U_j \partial_j T-\partial^2_{ll}T = 0 \text{ in } \mathbb{F}, \\
    (Pe_l U_j T-\partial_jT)n_j = 1  \text{ on } \mathbb{U},\\
    (Pe_l U_j T-\partial_jT)n_j = 0  \text{ on } \mathbb{D},\\
    T = 0   \text{ on } \partial\mathbb{M}, \\
    T \quad \text{periodic along $\bm{t}$,$\bm{s}$},
    \end{cases}
    \begin{cases}
    Pe_l U_j \partial_j S-\partial^2_{ll} S = 0  \text{ in } \mathbb{F},\\
    (Pe_l U_j S-\partial_j S)n_j = 0 \text{ on } \mathbb{U}, \\
    (Pe_l U_j S-\partial_j S)n_j = 1 \text{ on } \mathbb{D}, \\
    S = 0  \text{ on } \partial\mathbb{M}, \\ 
    S \quad \text{periodic along $\bm{t}$,$\bm{s}$}.
    \end{cases}
\label{eq:solvcondBFbeforeDirac}\end{equation}
The solvability conditions \eqref{eq:solvcondBFbeforeDirac} slightly differ from those introduced by \cite{zampogna2022transport} because of the presence of the advective term. In the case of flows on rough, impermeable surfaces, \cite{bottaroperspec} and \cite{lacis_sudhakar_pasche_bagheri_2020}  showed that systems \eqref{eq:solvcondBFbeforeDirac} are equivalent to the following set of equations
\begin{equation*}
    \begin{cases}
        Re_l U_m \partial_m M_{ij}=-\partial_i Q_{j} +\partial^2_{ll}M_{ij} +\delta_{\mathbb{C}}\delta_{ij},\\
        \partial_i M_{ij}=0,\\
        \Sigma_{pq}(M_{\cdot j},Q_{j})n_q=0 \quad \text{on $\mathbb{U}$, $\mathbb{D}$},\\
        M_{ij}=0 \quad \text{on $\partial \mathbb{M}$},\\
        M_{ij},Q_{j} \quad \text{periodic along $\bm{t}$,$\bm{s}$},
    \end{cases}
    \begin{cases}
        Re_l U_m \partial_m N_{ij}=-\partial_i R_{j} +\partial^2_{ll}N_{ij}-\delta_{\mathbb{C}}\delta_{ij},\\
        \partial_i N_{ij}=0,\\
        \Sigma_{pq}(N_{\cdot j},R_{j})n_q=0 \quad \text{on $\mathbb{U}$, $\mathbb{D}$},\\
        N_{ij}=0 \quad \text{on $\partial \mathbb{M}$},\\
        N_{ij},R_{j} \quad \text{periodic along $\bm{t}$,$\bm{s}$},
    \end{cases}
\end{equation*}
\begin{equation}
    \begin{cases}
    Pe_l U_j \partial_j T-\partial^2_{ll}T+\delta_{\mathbb{C}} = 0 \text{ in } \mathbb{F}, \\
    (Pe_l U_j T-\partial_jT)n_j = 0  \text{ on } \mathbb{U}, \mathbb{D},\\
    T = 0   \text{ on } \partial\mathbb{M}, \\
    T \quad \text{periodic along $\bm{t}$,$\bm{s}$},
    \end{cases}
    \begin{cases}
    Pe_l U_j \partial_j S-\partial^2_{ll} S-\delta_{\mathbb{C}} = 0  \text{ in } \mathbb{F},\\
    (Pe_l U_j S-\partial_j S)n_j = 0 \text{ on } \mathbb{U}, \mathbb{D}, \\
    S = 0  \text{ on } \partial\mathbb{M}, \\ 
    S \quad \text{periodic along $\bm{t}$,$\bm{s}$},
    \end{cases}
\label{eq:solvconddirac}\end{equation}
where $\delta_{\mathbb{C}}$ is a Dirac impulse centered in $x_{\mathbb{C}}$. The quantities $M_{ij},N_{ij},Q_i,R_i,T$ and $S$ depend parametrically on the closure advective velocity $U_j$. In the present paper we adopt formulation \eqref{eq:solvconddirac} to compute the microscopic solution.
\subsection{Averaging step and macroscopic condition}\label{sub:avgandmacro}
To upscale the microscopic solutions, we introduce the following averages at both the upstream and downstream sides of the membrane,
\begin{align}
    \bar{M}_{ij}=\frac{1}{|\mathbb{U}|}\int_{\mathbb{U}} M_{ij} dx_t dx_s, & \qquad \bar{N}_{ij}=\frac{1} {|\mathbb{D}|}\int_{\mathbb{D}} N_{ij} dx_t dx_s,\\
    \bar{T}=\frac{1}{|\mathbb{U}|}\int_{\mathbb{U}} T dx_t dx_s, & \qquad \bar{S}=\frac{1} {|\mathbb{D}|}\int_{\mathbb{D}} S dx_t dx_s. 
\label{eq:avgcentral}\end{align}
Applying averages \eqref{eq:avgcentral} to equation \eqref{eq:model}, The following macroscopic boundary conditions are obtained,
\begin{equation}
    \begin{cases}
        \bar{u}_i^{\mathbb{O}}=\epsilon Re_L(\bar{M}_{ij}\Sigma_{jk}^{\mathbb{O,U}}n_k+\bar{N}_{ij}\Sigma_{jk}^{\mathbb{O,D}}n_k) \\
                \bar{c}^{\mathbb{O}}=\epsilon (\bar{T} F_j^{\mathbb{O,U}}n_j+\bar{S} F_j^{\mathbb{O,D}}n_j).
    \end{cases}
\label{eq:macromodel}\end{equation}
As thoroughly explained in \cite{zampogna2020effective}, $\bar{M}_{ij},\bar{N}_{ij}, \bar{T}$ and $\bar{S}$ have a precise physical meaning. In opposition to \cite{zampogna2020effective, zampogna2022transport}, $\bm{M},\bm{N}, S, T$ are not solely properties of the geometry, but also of the fluid flow since they depend on the closure advective velocity. $\bar{M}_{nn}$ and $\bar{M}_{tt}$ represent a permeability and a slip coefficient, respectively. In the case of a flow around an anisotropic (i.e. asymmetric about the normal axis) inclusion, the normal or tangential forcing causes an average fluid flow which is both normal and tangential. This manifests as non-zero values of the off-diagonal components of $\bar{M}_{ij}, \bar{N}_{ij}$.  
\subsection{The closure advective velocity}\label{sub:lineariz}
The solution to problem \eqref{eq:solvconddirac} lies in the closure advective velocity that gives rise to an inertia-driven coupling between the macroscopic and microscopic flows. Two different definitions have been considered in the present work:
\begin{enumerate}
\item a constant advective velocity in the microscopic cell
\begin{equation}
    U_i=\bar{u}_i^{\mathbb{O}},
\label{eq:Uiconstant advection closure}\end{equation}
which leads to an Oseen-like equation (constant advection closure). This approach has already been proposed in the volume-averaged and homogenization frameworks by other authors for the flow over rough surfaces \citep{zampogna_slip,bottaroperspec} and the flow in bulk porous media \citep{valdesparadaFlowNear2021,sánchez-vargas_valdés-parada2023};
\item a spatially-dependent closure advective velocity, reconstructed from the outer stresses (variable advection closure),
\begin{equation}
    U_i=\epsilon Re_L \left( M_{ij}\Sigma_{jk}^{\mathbb{O,U}}n_k+N_{ij}\Sigma_{jk}^{\mathbb{O,D}}n_k \right).
\label{eq:Uivariable advection closure}\end{equation}
\end{enumerate}
The equations for $M_{ij}, Q, T$ in the microscopic problems thus become
\begin{itemize}
    \item for the constant advection closure approach \eqref{eq:Uiconstant advection closure}:
    \begin{equation*}
    \begin{cases}
        \epsilon Re_L \frac{\mathcal{U}}{\mathcal{U}^{\mathbb{O}}}\bar{u}_m^{\mathbb{O}} \partial_m M_{ij}=-\partial_i Q_{j} +\partial^2_{ll}M_{ij} +\delta_{\mathbb{C}}\delta_{ij},\\
        \partial_i M_{ij}=0,\\
        \Sigma_{pq}(M_{\cdot j},Q_{j})n_q=0 \quad \text{on $\mathbb{U}$, $\mathbb{D}$},\\
        M_{ij}=0 \quad \text{on $\partial \mathbb{M}$},\\
        M_{ij},Q_{j} \quad \text{periodic along $\bm{t}$,$\bm{s}$};
    \end{cases}
    \begin{cases}
        \epsilon Re_L \frac{\mathcal{U}}{\mathcal{U}^{\mathbb{O}}}\bar{u}_m^{\mathbb{O}} \partial_m N_{ij}=-\partial_i R_{j} +\partial^2_{ll}N_{ij} -\delta_{\mathbb{C}}\delta_{ij},\\
        \partial_i N_{ij}=0,\\
        \Sigma_{pq}(N_{\cdot j},R_{j})n_q=0 \quad \text{on $\mathbb{U}$, $\mathbb{D}$},\\
        N_{ij}=0 \quad \text{on $\partial \mathbb{M}$},\\
        N_{ij},R_{j} \quad \text{periodic along $\bm{t}$,$\bm{s}$};
    \end{cases}
    \end{equation*}
    
    \begin{equation}
    \hskip-15pt
    \begin{cases}
        \epsilon Pe_L \frac{\mathcal{U}}{\mathcal{U}^{\mathbb{O}}}\bar{u}_i^{\mathbb{O}} \partial_j T-\partial^2 T _{ll}+\delta_{\mathbb{C}} = 0, \\
        (\epsilon Pe_L \bar{u}_i^{\mathbb{O}} T-\partial_jT)n_j = 0  \text{ on } \mathbb{U}, \mathbb{D},\\
        T = 0   \text{ on } \partial\mathbb{M}, \\
        T \quad \text{periodic along $\bm{t}$,$\bm{s}$};
    \end{cases}
    \begin{cases}
        \epsilon Pe_L \frac{\mathcal{U}}{\mathcal{U}^{\mathbb{O}}}\bar{u}_i^{\mathbb{O}} \partial_j S-\partial^2 S _{ll}-\delta_{\mathbb{C}} = 0,\\
        (\epsilon Pe_L \bar{u}_i^{\mathbb{O}} S-\partial_j S)n_j = 0 \text{ on } \mathbb{U}, \mathbb{D}, \\
        S = 0  \text{ on } \partial\mathbb{M}, \\ 
        S \quad \text{periodic along $\bm{t}$,$\bm{s}$};
    \end{cases}
    \label{eq:solvcondCAC}\end{equation}
    \item for the variable advection closure approach \eqref{eq:Uivariable advection closure}:
    \begin{equation*}
    \begin{cases}
        \epsilon^2 Re_L^2 \frac{\mathcal{U}}{\mathcal{U}^{\mathbb{O}}}(M_{mn}\Sigma_{nl}^{\mathbb{O,U}}n_l+N_{mn}\Sigma_{nl}^{\mathbb{O,D}}n_l) \partial_m M_{ij}=-\partial_i Q_{j} +\partial^2_{ll}M_{ij} +\delta_{\mathbb{C}}\delta_{ij},\\
        \partial_i M_{ij}=0,\\
        \Sigma_{pq}(M_{\cdot j},Q_{j})n_q=0 \quad \text{on $\mathbb{U}$, $\mathbb{D}$},\\
        M_{ij}=0 \quad \text{on $\partial \mathbb{M}$},\\
        M_{ij},Q_{jk} \quad \text{periodic along $\bm{t}$,$\bm{s}$};
    \end{cases}
    \end{equation*}
    \begin{equation*}
    \begin{cases}
        \epsilon^2 Re_L^2 \frac{\mathcal{U}}{\mathcal{U}^{\mathbb{O}}}(M_{mn}\Sigma_{nl}^{\mathbb{O,U}}n_l+N_{mn}\Sigma_{nl}^{\mathbb{O,D}}n_l) \partial_m N_{ij}=-\partial_i R_{j} +\partial^2_{ll}N_{ij} -\delta_{\mathbb{C}}\delta_{ij},\\
        \partial_i N_{ij}=0,\\
        \Sigma_{pq}(N_{\cdot j},R_{j})n_q=0 \quad \text{on $\mathbb{U}$, $\mathbb{D}$},\\
        N_{ij}=0 \quad \text{on $\partial \mathbb{M}$},\\
        N_{ij},R_{j} \quad \text{periodic along $\bm{t}$,$\bm{s}$};
    \end{cases}
    \end{equation*}
    \begin{equation*}
    \begin{cases}
        \epsilon^2 Pe_L \frac{\mathcal{U}}{\mathcal{U}^{\mathbb{O}}}(M_{mn}\Sigma_{nl}^{\mathbb{O,U}}n_l+N_{mn}\Sigma_{nl}^{\mathbb{O,D}}n_l) \partial_j T-\partial^2_{ll} T+\delta_{\mathbb{C}} = 0, \\
        (\epsilon^2 Pe_L (M_{mn}\Sigma_{nl}^{\mathbb{O,U}}n_l+N_{mn}\Sigma_{nl}^{\mathbb{O,D}}n_l) T-\partial_jT)n_j = 0  \text{ on } \mathbb{U}, \mathbb{D},\\
        T = 0   \text{ on } \partial\mathbb{M}, \\
        T \quad \text{periodic along $\bm{t}$,$\bm{s}$};
    \end{cases}
    \end{equation*}
    \begin{equation}
    \begin{cases}
        \epsilon^2 Pe_L \frac{\mathcal{U}}{\mathcal{U}^{\mathbb{O}}}(M_{mn}\Sigma_{nl}^{\mathbb{O,U}}n_l+N_{mn}\Sigma_{nl}^{\mathbb{O,D}}n_l) \partial_j S-\partial^2_{ll} S-\delta_{\mathbb{C}} = 0,\\
        (\epsilon^2 Pe_L (M_{mn}\Sigma_{nl}^{\mathbb{O,U}}n_l+N_{mn}\Sigma_{nl}^{\mathbb{O,D}}n_l) S-\partial_j S)n_j = 0 \text{ on } \mathbb{U}, \mathbb{D}, \\
        S = 0  \text{ on } \partial\mathbb{M}, \\ 
        S \quad \text{periodic along $\bm{t}$,$\bm{s}$}.
    \end{cases}
    \label{eq:solvcondVAC}\end{equation}
\end{itemize}
The tensors and scalars used in equation \eqref{eq:macromodel} are found either by solving problem \eqref{eq:solvcondCAC} or \eqref{eq:solvcondVAC}. A computational iterative strategy to interface the macroscopic fields with the microscopic problems is required.

%% file: 3A-micro_results.tex
\section{Solution of the microscopic problems}

\label{sec:microfields} 
We investigate the influence of the closure advective velocity on the microscopic fields $M_{ij}, N_{ij}, T$ and $S$, in the constant and variable advection closure cases. We introduce the porosity $\theta=|\mathbb{C_F}|/|\mathbb{C_F}\cup\mathbb{C_M}|$ as the fluid-to-total ratio at the membrane centerline $\mathbb{C}$ (cf. figure\ref{fig:generaldomain}a). As a benchmark, we consider a circular inclusion of porosity $\theta=0.7$ (cf. figure \ref{fig:ose_fields}$a$). The solution of equations \eqref{eq:solvconddirac} is computed numerically using the finite-element software COMSOL Multiphysics 6.0. We refer to appendix \ref{app:computat_details} for further details about the numerical solution.
\subsection{Constant advection closure}
\label{sec:constant advection closure_micro}
\begin{figure}
    \centering
    \includegraphics[width=\textwidth]{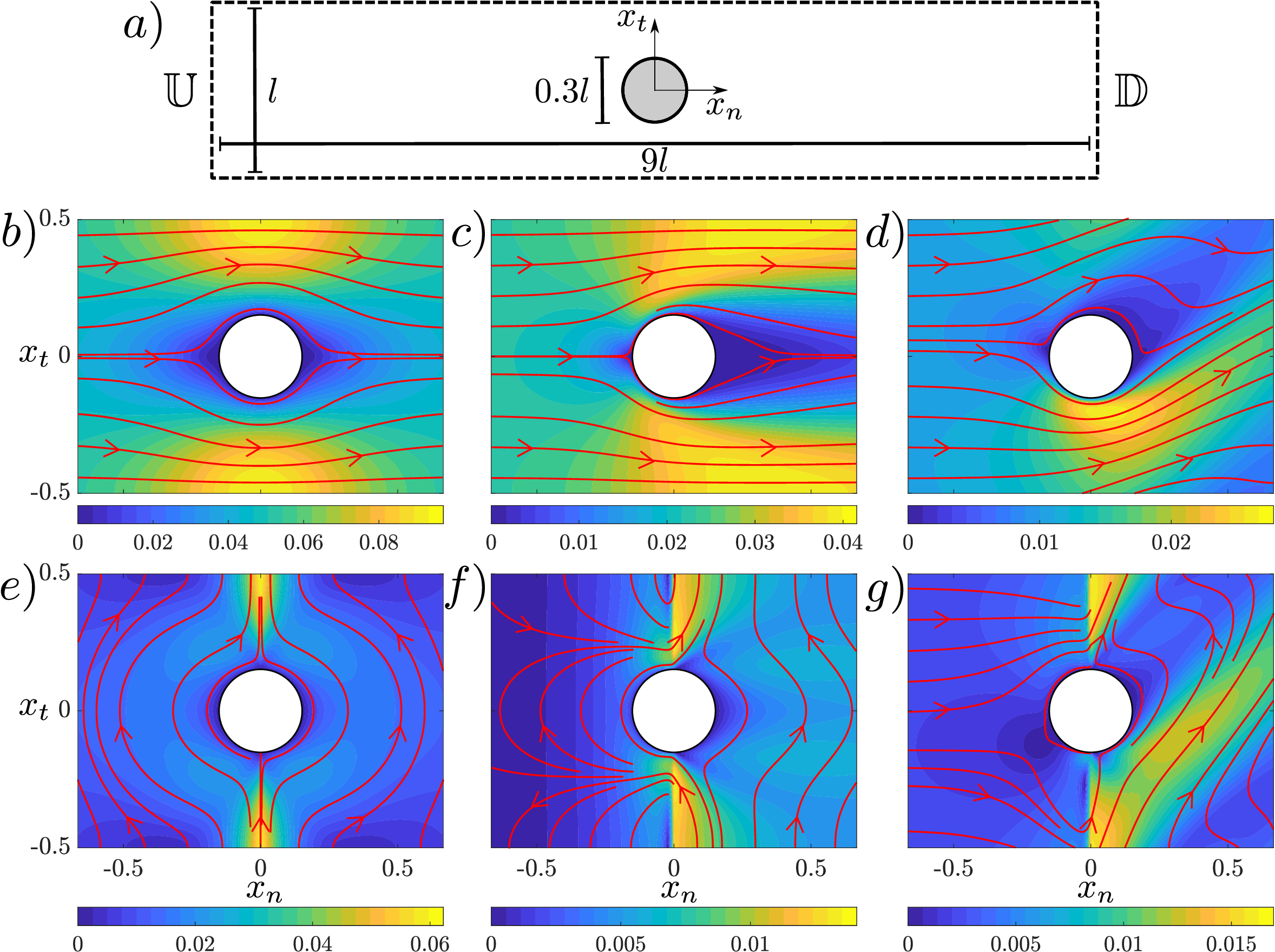}
    \caption{Panel $a$: microscopic domain. ($b-g$) Magnitude contours and streamlines (red) of $(M_{nn},M_{tn})$ for $\check{u}_n=0,\check{u}_t=0$ (panel $b$), $\check{u}_n=50,\check{u}_t=0$ ($c$), $\check{u}_n=50,\check{u}_t=50$ ($d$). Magnitude contours and streamlines of the tensors $(M_{nt},M_{tt})$ for $\check{u}_n=0,\check{u}_t=0$ (panel $e$), $\check{u}_n=50,\check{u}_t=0$ ($f$), $\check{u}_n=50,\check{u}_t=50$ ($g$). Fields are obtained using the constant advection closure approach.}
    \label{fig:ose_fields}
\end{figure}

In the constant advection closure problem, we specify the advective velocity in \eqref{eq:solvconddirac} as a constant field \eqref{eq:Uiconstant advection closure}, obtaining \eqref{eq:solvcondCAC}. The solutions for the couples $(M_{nn},M_{tn}),(M_{nt},M_{tt})$ are presented in figure \ref{fig:ose_fields}. To compact the notation, we introduce $\check{u}_i=\epsilon Re_L \frac{\mathcal{U}}{\mathcal{U}^{\mathbb{O}}}\bar{u}_i^{\mathbb{O}}$ in the term in front of the convective term in the LHS of equations \eqref{eq:solvcondCAC}, the inertia-driven coupling term with the macroscopic problem. We consider three values of $\check{u}_i$, corresponding to the pure diffusive case  ($\check{u}_i=0$, frames $b,e$), a case of pure normal advection ($\check{u}_n\neq 0$ and $\check{u}_t=0$, frames $c,f$) and a case of normal and tangential advection ($\check{u}_n\neq 0$, $\check{u}_t\neq 0$, frames $d,g$). Recirculating zones propagate downstream the inclusion in the direction of the advective flow. Note that the same microscopic behaviour is noticed for $N_{ij}$, which satisfies $N_{ij}(\check{u}_i)=-M_{ij}(-\check{u}_i)$, and it is hence not shown. 

\begin{figure}
    \centering
    \includegraphics[width=\textwidth]{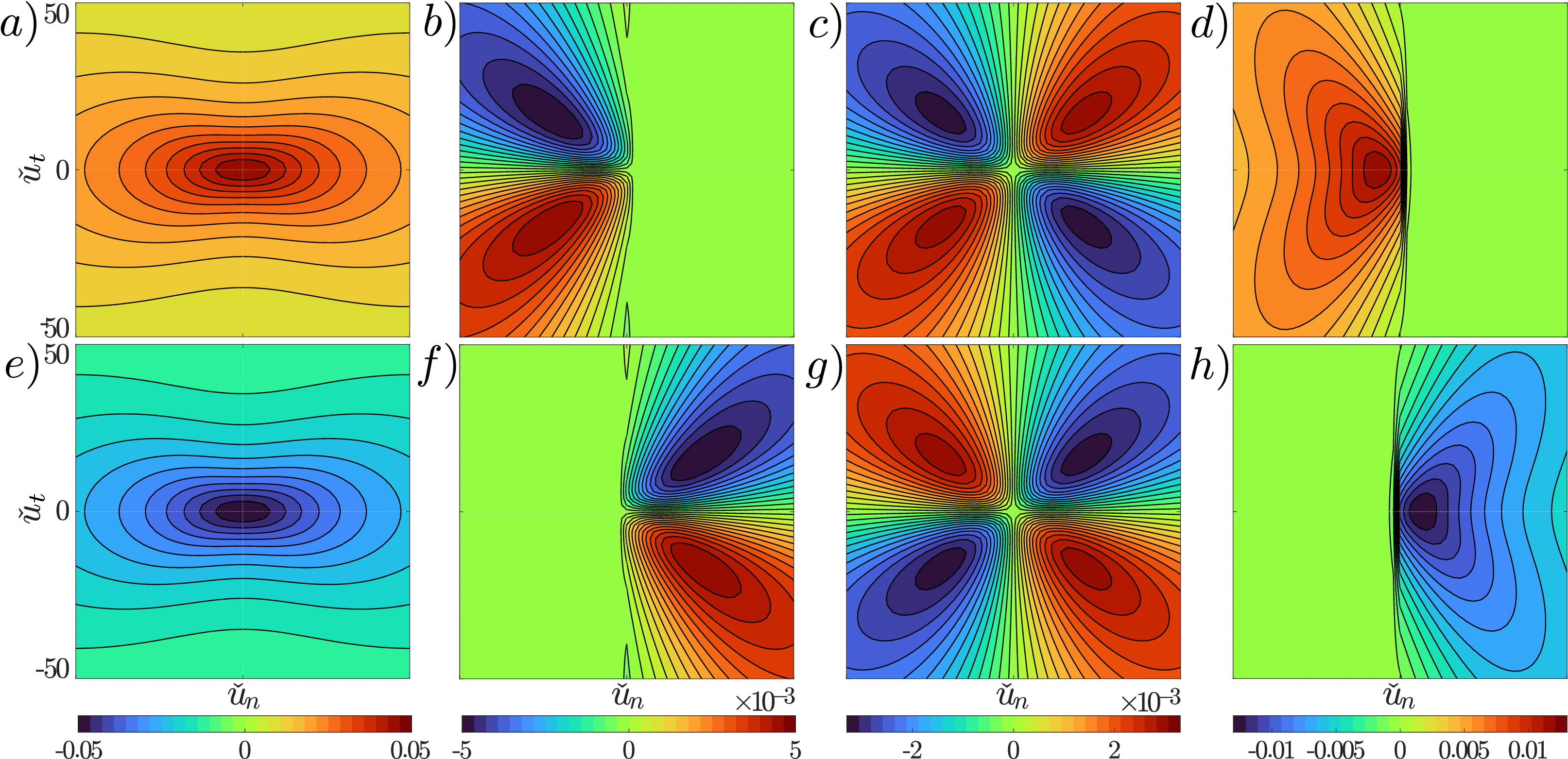}
    \caption{Average tensor components using the constant advection approach: $\bar{M}_{nn}$ (panel $a$), $\bar{M}_{nt}$ ($b$), $\bar{M}_{tn}$ ($c$), $\bar{M}_{tt}$ ($d$), $\bar{N}_{nn}$ ($e$), $\bar{N}_{nt}$ ($f$), $\bar{N}_{tn}$ ($g$), $\bar{N}_{tt}$ ($h$).}
    \label{fig:ose_cir_contourf}
\end{figure}

By applying the averaging operators \eqref{eq:avgcentral} to these fields for $\check{u}_i$ ranging in $[-50,50]$ we obtain the maps of $\bar{M}_{ij}, \bar{N}_{ij}$. Figure \ref{fig:ose_cir_contourf} shows the contours of $\bar{M}_{ij}$ and $\bar{N}_{ij}$ as functions of $\check{u}_i$. The off-diagonal components of $\bar{M}_{ij},\bar{N}_{ij}$ are zero when $\check{u}_i=0$. The $\cdot_{nt}$ and $\cdot_{tt}$ components show a strong asymmetry with respect to $\check{u}_n$. The permeability $\bar{M}_{nn}$ (figure \ref{fig:ose_cir_contourf}a) is instead symmetric with respect to both components of $\check{u}_i$ and shows a maximum for $\check{u}_i=0$. This suggests that inertia always decreases the permeability unless both the off-diagonal components are non-zero and partially compensate for the diminished $\bar{M}_{nn}$. Similar considerations apply to $\bar{N}_{ij}$, since $\bar{M}_{ij}(\check{u}_k)=-\bar{N}(-\check{u}_k)$.

\begin{figure}
    \centering
    \includegraphics[width=\textwidth]{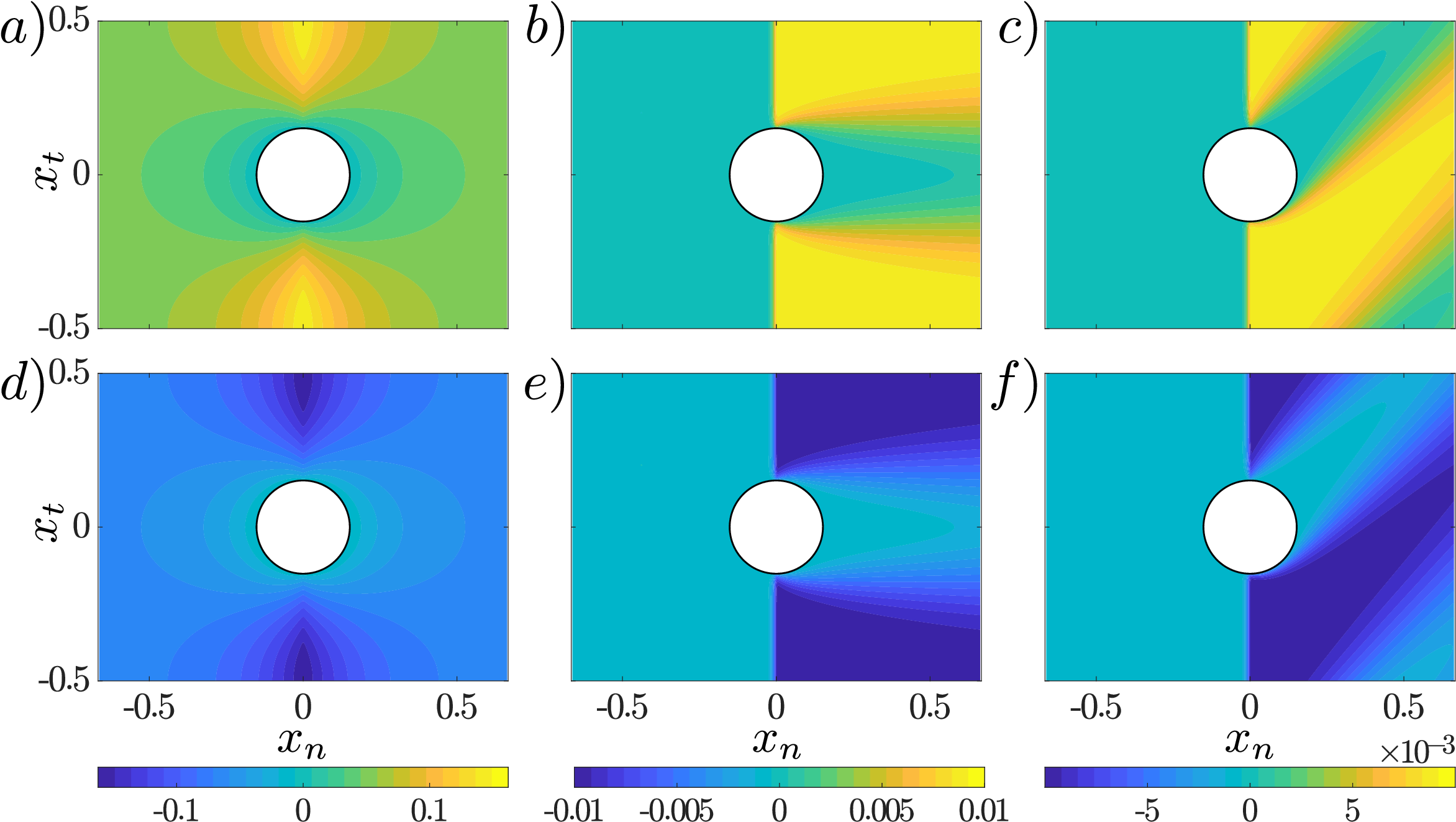}
    \caption{Contours of $T$ (panels $a-c$) and $S$ (panels $d-f$) for $\tilde{u}_n=\tilde{u}_t=0$ ($a,d$), $\tilde{u}_n=100, \tilde{u}_t=0$ ($b,e$) and $\tilde{u}_n=\tilde{u}_t=100$ ($c,f$) obtained using the constant advection closure approach.}
    \label{fig:ose_Tmapfields}
\end{figure}

We consider now the problem for $T$ and $S$. We parameterize $T$ and $S$ in terms of the quantities appearing in the advective term, compacted as $\tilde{u}_i=\epsilon Pe_L \frac{\mathcal{U}}{\mathcal{U}^{\mathbb{O}}} \bar{u}_i^{\mathbb{O}}$. Figure \ref{fig:ose_Tmapfields} shows $T$ and $S$ in the pure diffusive case (panel $a$), an advective case with $\tilde{u}_n=100, \tilde{u}_t=0$ (panel $b$) and a case with $\tilde{u}_n=\tilde{u}_t=100$. 

\begin{figure}
    \centering
    \includegraphics[width=\textwidth]{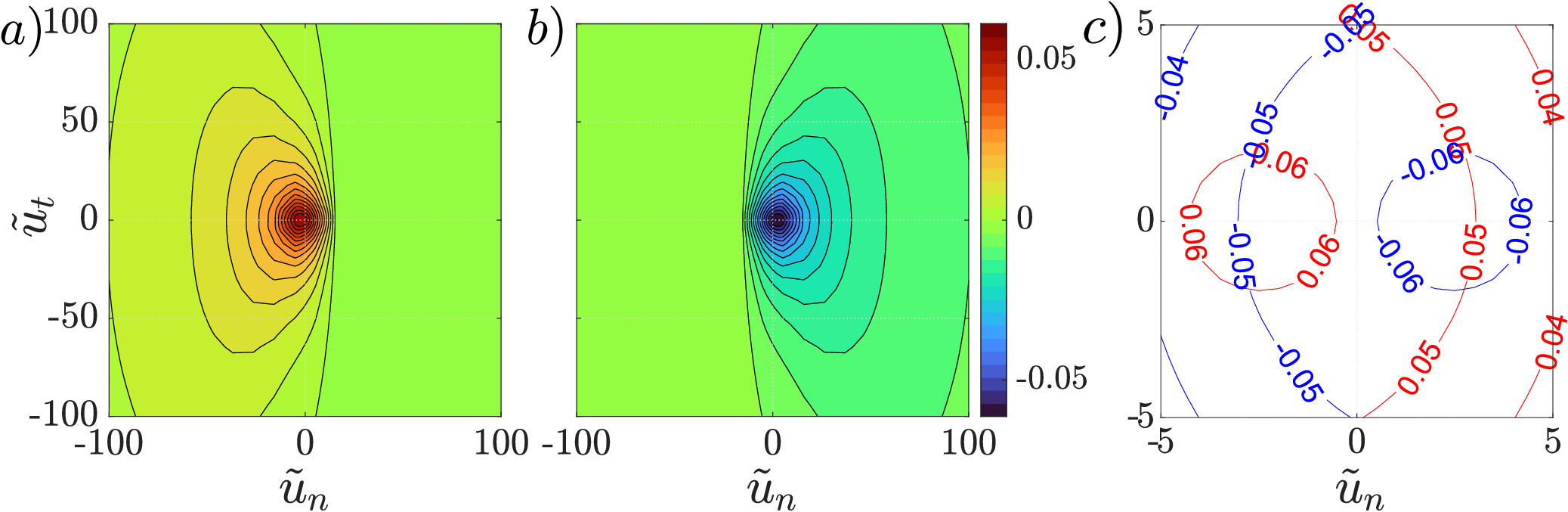}
    \caption{Contours of $\bar{T}$ (panel $a$) and $\bar{S}$ ($b$) as a function of the advection velocity components $\tilde{u}_n,\tilde{u}_t$ obtained using the constant advection closure approach. ($c$) A zoom-in on the region where the maximum and minimum of $\bar{T}$ (blue) and $\bar{S}$ (red) are attained.}
    \label{fig:ose_TSmaps}
\end{figure}

Maps of $\bar{T}$ are obtained by averaging $T$ using equation \eqref{eq:avgcentral} for $\tilde{u}_i\in [-100,100]$ (figure \ref{fig:ose_TSmaps}). We notice that $\bar{T}(\tilde{u}_i,\tilde{u}_t)=\bar{T}(\tilde{u}_n,-\tilde{u}_t)$ and that the maximum (minimum) of $\bar{T}$ ($\bar{S}$) is attained for a non-zero $\tilde{u}_n$. This suggests that advection increases the effective diffusivity $\bar{T}$. Similar considerations apply for $S$ fields, which obey $\bar{S}(\tilde{u}_i)=-\bar{T}(-\tilde{u}_i)$.
Eventually, the advective velocity can cause recirculating zones or concentration wakes downstream of the inclusions, with non-zero off-diagonal components of the tensors even in the absence of geometrical asymmetry (cf. panels \ref{fig:ose_cir_contourf}b,c,f,g). 

\subsection{Variable advection closure}
\label{sec:variable advection closure_micro}
\begin{figure}
    \centering
    \includegraphics[width=\textwidth]{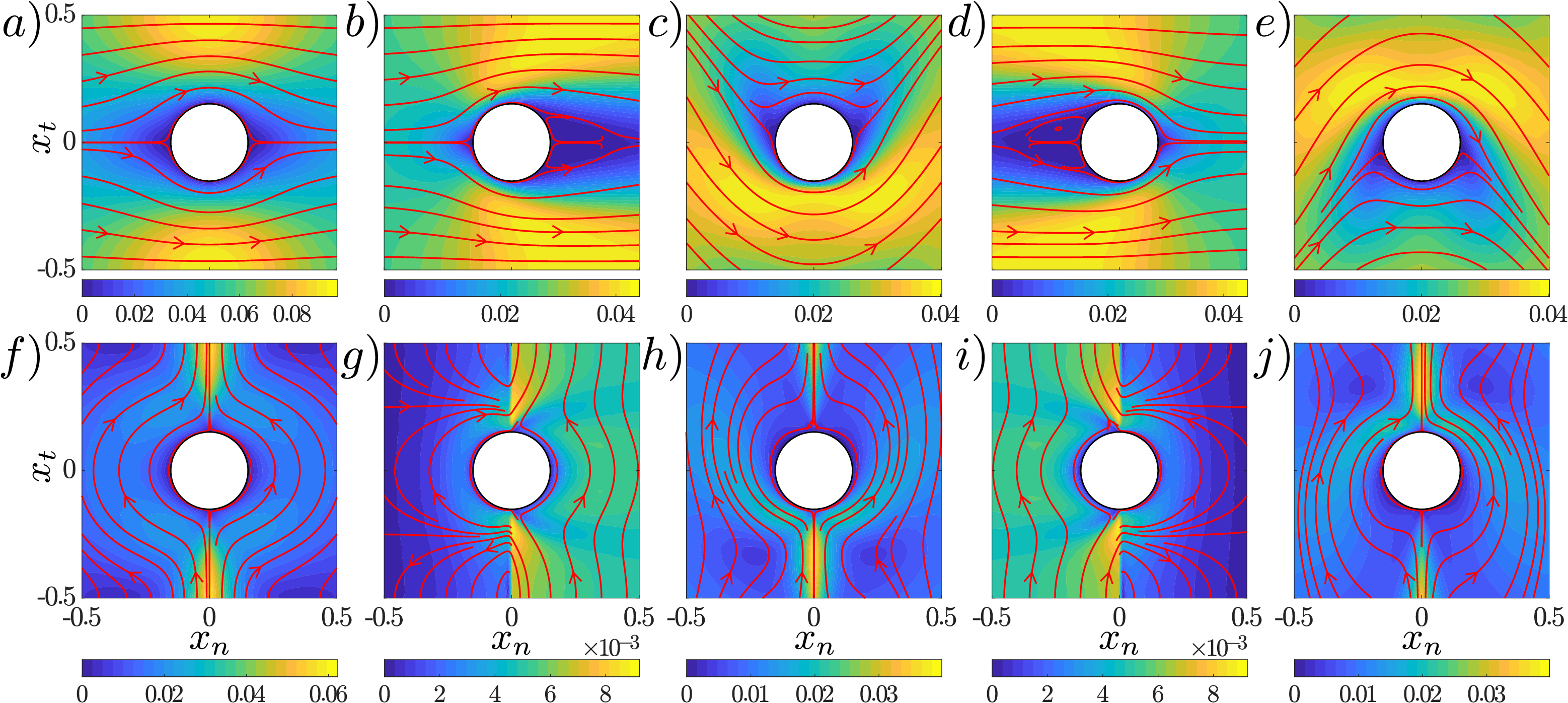}
    \caption{Magnitude iso-contours and streamlines (red) of $(M_{nn},M_{tn})$ (panels $a-e$) and $(M_{nt},M_{tt})$ (panels $f-j$) for 5 different combinations of $\check{\Sigma}^{\mathbb{U},\mathbb{D}}_{ij}$. $\check{\Sigma}^{\mathbb{U},\mathbb{D}}_{ij}=\epsilon^2Re_L^2\Sigma^{\mathbb{U},\mathbb{D}}_{ij}=0$ (panels $a,f$), $\check{\Sigma}^{\mathbb{U}}_{nn}=2500$ ($b,g$), $\check{\Sigma}^{\mathbb{U}}_{tn}=2500$ ($c,h$), $\check{\Sigma}^{\mathbb{D}}_{nn}=2500$ ($d,i$) and $\check{\Sigma}^{\mathbb{D}}_{tn}=2500$ ($e,j$). For each case, the components of $\check{\Sigma}^{\mathbb{U,D}}_{ij}$ not specified are equal to zero. The variable advection closure is employed.}
    \label{fig:NSE_momentum_FIELDS}
\end{figure}

\begin{figure}
    \centering
    \includegraphics[width=\textwidth]{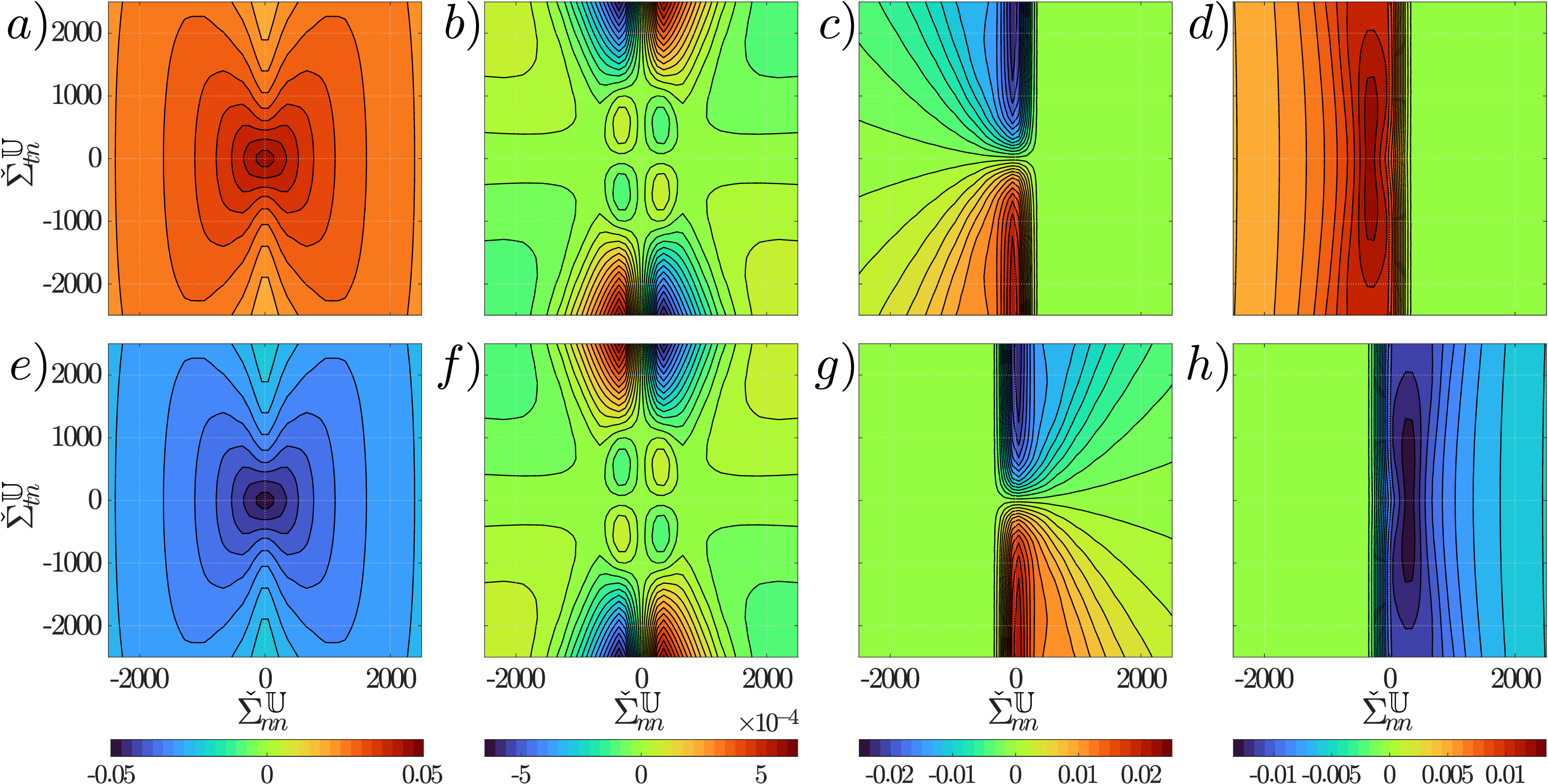}
    \caption{Iso-contours of average tensor component $\bar{M}_{nn}$ (panel $a$), $\bar{M}_{nt}$ ($b$), $\bar{M}_{tn}$ ($c$), $\bar{M}_{tt}$ ($d$), $\bar{N}_{nn}$ ($e$),  $\bar{N}_{nt}$ ($f$), $\bar{N}_{tn}$ ($g$), $\bar{N}_{tt}$ ($h$) for $(\check{\Sigma}^{\mathbb{U}}_{nn},\check{\Sigma}^{\mathbb{U}}_{tn})\in[-2500,2500]$, while $\check{\Sigma}^{\mathbb{D}}_{\cdot n}=0$. The variable advection closure approach is exploited.}
    \label{fig:NSE_momentum_MAPS}
\end{figure}

In the variable advection closure problem, we specify the advective velocity in \eqref{eq:solvconddirac} as a variable field, see equation \eqref{eq:Uivariable advection closure}, obtaining \eqref{eq:solvcondVAC}. The solutions for $M_{ij}$ and $N_{ij}$ are presented in figure \ref{fig:ose_fields}. The advective term depends on four parameters, $\check{\Sigma}^{\mathbb{U},\mathbb{D}}_{ij}=\epsilon^2Re_L^2 \frac{\mathcal{U}}{\mathcal{U}^{\mathbb{O}}} \Sigma^{\mathbb{U},\mathbb{D}}_{ij}n_j$ for the hydrodynamic problem and four parameters $\tilde{\Sigma}^{\mathbb{U},\mathbb{D}}_{ij}=\epsilon^2 Pe_L \frac{\mathcal{U}}{\mathcal{U}^{\mathbb{O}}} \Sigma^{\mathbb{U},\mathbb{D}}_{ij}n_j$ for the advection-diffusion problem ($T,S$). In figure \ref{fig:NSE_momentum_FIELDS} the effect of $\check{\Sigma}_{ij}^{\mathbb{U,D}}$ is shown for variations of some sample components of $M_{ij}$. 
The application of a non-zero stress component along a given direction causes the flow to deviate along that direction, eventually developing a laminar separation bubble downstream (for $\check{\Sigma}_{nn}$ with the same sign of the Dirac forcing) or upstream the inclusion (for $\check{\Sigma}_{nn}$ with opposed sign with respect to the Dirac forcing). 
By applying the averaging operators \eqref{eq:avgcentral} to $M_{ij}$ and $N_{ij}$, the iso-levels of $\bar{M}_{ij},\bar{N}_{ij}$ are obtained for varying $\check{\Sigma}_{ij}^{\mathbb{U,D}}$ (cf. figure \ref{fig:NSE_momentum_MAPS}). This figure represents a sampling on a 2D sub-manifold of the four-dimensional manifold where the averaged tensors live. Further sub-manifolds are presented in appendix \ref{app:maps} for other values of $\check{\Sigma}_{ij}^{\mathbb{U,D}}$. The $\bar{M}_{nn}$ and $\bar{N}_{nn}$ terms show symmetry about two axes also in this case, while $\bar{M}_{ij}(\check{\Sigma}^{\mathbb{U},\mathbb{D}}_{ij})=-N(-\check{\Sigma}^{\mathbb{U},\mathbb{D}}_{ij})$. The maxima of permeability ($\bar{M}_{nn}$) are found in the case of Stokes flow (i.e. balanced $\check{\Sigma}_{ij}^{\mathbb{U}}$ and $\check{\Sigma}_{ij}^{\mathbb{D}}$ contributions). Slip ($\bar{M}_{tt}$) has a maximum for values of $\check{\Sigma}^{\mathbb{U}}_{nn}$ close to $\check{\Sigma}^{\mathbb{D}}_{nn}$, but not exactly corresponding to Stokes' flow.

\begin{figure}
    \centering
    \includegraphics[width=\textwidth]{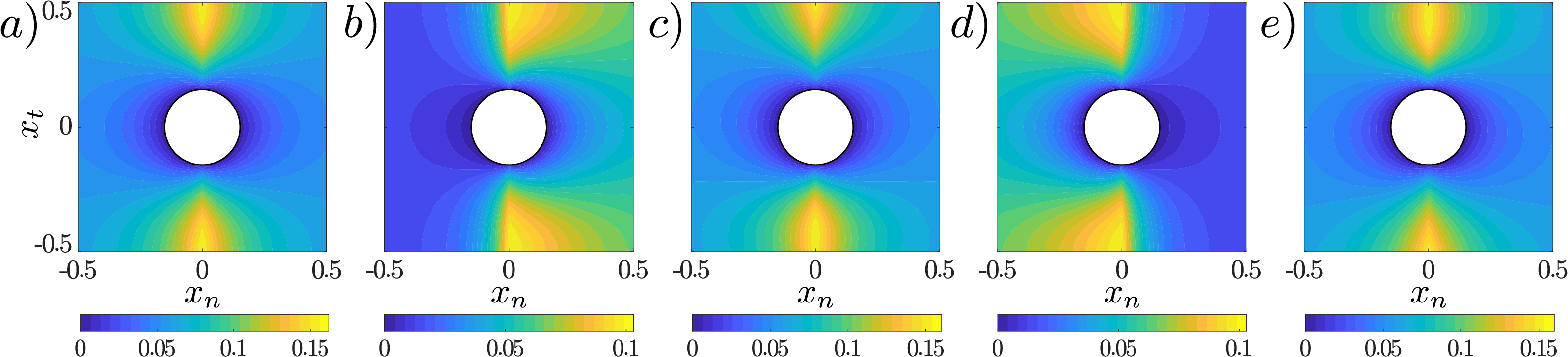}
    \caption{Fields of the effective diffusivity $T$ computed using the variable advection closure model. $\tilde{\Sigma}^{\mathbb{U},\mathbb{D}}_{ij}=0$ (panel $a$), $\tilde{\Sigma}^{\mathbb{U}}_{nn}=100$ ($b$), $\tilde{\Sigma}^{\mathbb{U}}_{tn}=100$ ($c$), $\tilde{\Sigma}^{\mathbb{D}}_{nn}=100$ ($d$) and $\tilde{\Sigma}^{\mathbb{D}}_{tn}=100$ ($e$). For each panel, the components of $\tilde{\Sigma}_{ij}^{\mathbb{U,D}}$ not mentioned are equal to zero.}
    \label{fig:fields_ts_NSE}
\end{figure}

The variable advection approach applied to the problem for $T$ and $S$ in \eqref{eq:solvconddirac} gives equations \eqref{eq:solvcondVAC}. Its solution requires the tensors $M_{ij}$ and $N_{ij}$ to be known. For simplicity, we consider the case of diffusive momentum transport ($Re_L=0$, panel $a$, corresponding to $\bar{M}_{nn}=0.05$, $\bar{M}_{nn}=0.01$, $\bar{M}_{nt}=\bar{M}_{tn}=0$ and $\bar{N}_{ij}=-\bar{M}_{ij}$). The microscopic field $T$ is presented in figure \ref{fig:fields_ts_NSE} for different values of $\tilde{\Sigma}_{ij}^{\mathbb{U,D}}$. $T$ exhibits a wake directed as the dominant inertial component for each case. By applying the average operator \eqref{eq:avgcentral}, we obtain the contours of $\bar{T}$ for $\tilde{\Sigma}^{\mathbb{U},\mathbb{D}}_{ij}\in [-50,50]$. Interestingly, in the considered range, there is a negligible influence of $\tilde{\Sigma}^{\mathbb{U},\mathbb{D}}_{tn}$, while $\tilde{\Sigma}^{\mathbb{U},\mathbb{D}}_{nn}$ is dominant in this problem. Similar considerations apply to $\bar{S}$, since $\bar{T}(\tilde{\Sigma}^{\mathbb{U},\mathbb{D}}_{nn})=-\bar{S}(-\tilde{\Sigma}^{\mathbb{U},\mathbb{D}}_{nn})$.
\begin{figure}
    \centering
    \includegraphics[width=\textwidth]{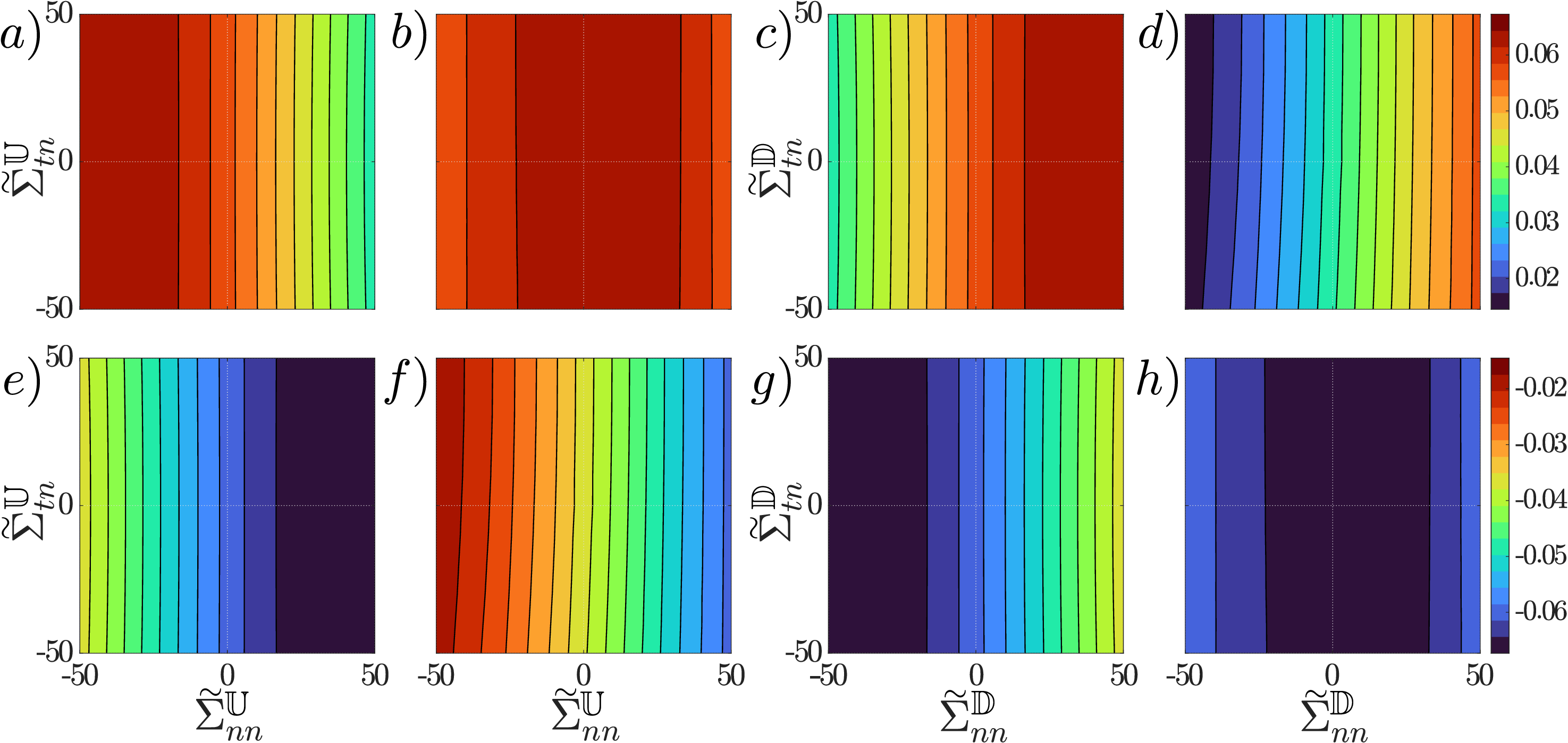}
    \caption{Contours of $\bar{T}$ (panels $a-d$) and $\bar{S}$ ($e-h$). Each panel is obtained for a different value of   $\check{\Sigma}^{\mathbb{U},\mathbb{D}}_{ij}$; $\tilde{\Sigma}^{\mathbb{D}}_{nn}=\tilde{\Sigma}^{\mathbb{D}}_{tn}=0$ (panels $a,e$), $\tilde{\Sigma}^{\mathbb{D}}_{nn}=\tilde{\Sigma}^{\mathbb{D}}_{tn}=50$ ($b,f$), $\tilde{\Sigma}^{\mathbb{U}}_{nn}=\tilde{\Sigma}^{\mathbb{U}}_{tn}=0$ ($c,g$), $\tilde{\Sigma}^{\mathbb{U}}_{nn}=\tilde{\Sigma}^{\mathbb{U}}_{tn}=50$ ($d,h$). The stress tensor components not mentioned in each panel are equal to zero. The variable advection closure approach is considered.}
    \label{fig:MAPS_ts_NSE}
\end{figure}

In this section, we presented relevant features of the microscopic flow occurring for non-negligible inertia. The microscopic solutions depend on the pore geometry and flow characteristics. A comparison of figures \ref{fig:ose_TSmaps} and \ref{fig:MAPS_ts_NSE} shows that $T$ depends strongly on on $\tilde{u}_t$ but not on $\tilde{\Sigma}_{tn}^{\mathbb{U,D}}$. However, this apparent discrepancy can be explained by considering that not all points in figure \ref{fig:ose_TSmaps} are images of points in figure \ref{fig:MAPS_ts_NSE} through equation \eqref{eq:macromodel}. The range represented in figure \ref{fig:MAPS_ts_NSE} thus corresponds only to a thin zone around the axis $\tilde{u}_t=0$ in figure \ref{fig:ose_TSmaps}.

%% file: 3B-macro_results.tex
\section{Comparison between full-scale simulations and homogenized model} \label{sec:fullscalemacro}
\begin{figure}
    \centering
    \includegraphics[width=0.95\textwidth]{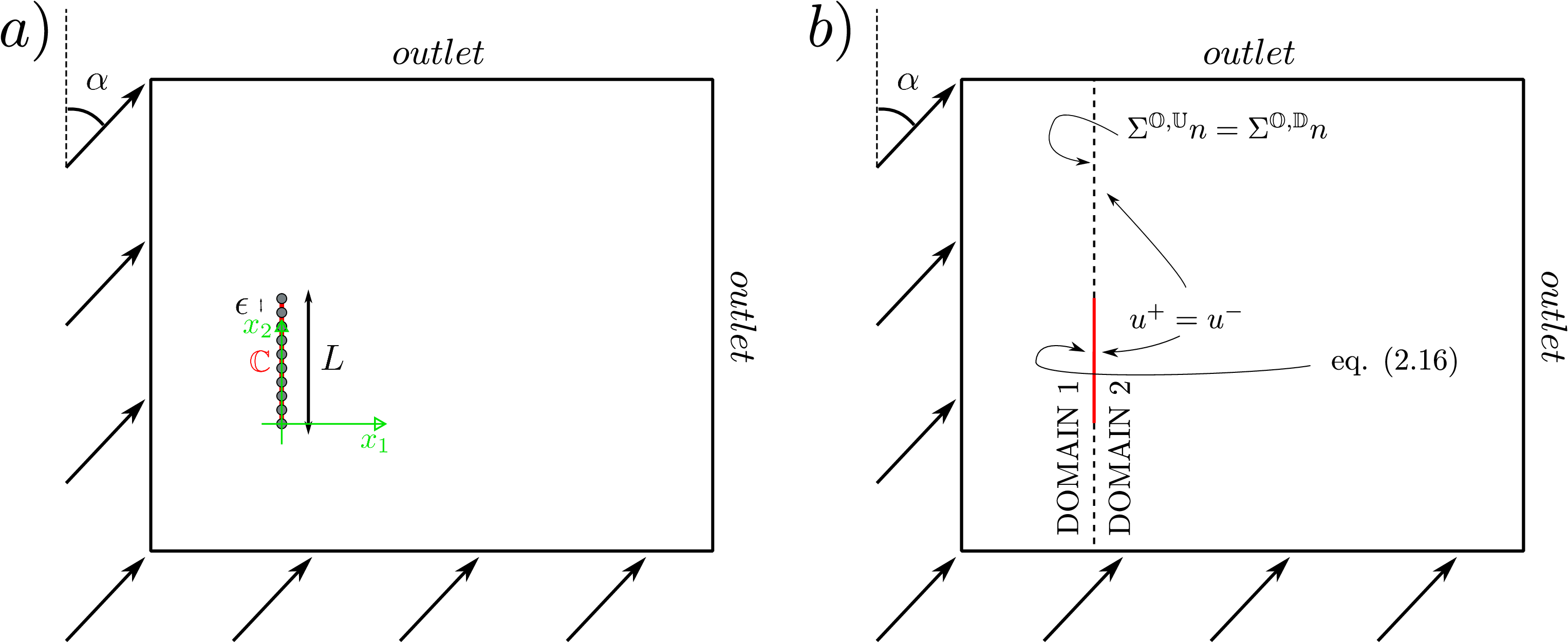}
    \caption{Panel $a)$: full-scale computational domain. Arrows correspond to a Dirichlet boundary condition on velocity (inlet), and outlet boundary condition corresponds to imposing $\Sigma_{ij} n_j=0$. No slip is applied on the membrane walls. The origin of the axes is placed at the lowest end of the membrane and the domain extends for $x_1\in[-1.5L,5.5L]$ and $x_2\in[-1.5L,3.5L]$. Panel $b)$: purely macroscopic computational domain. The full domain is separated into an upward and a downward domain by membrane centerline $\mathbb{C}$ (red).} 
    \label{fig:macrogeom}
\end{figure}

In this section, we compare the macroscopic solution against simulations of the flow solved at all scales. We consider a two-dimensional flat membrane composed of circular inclusions with spacing $l/L=\epsilon=0.1$, invested by a uniform stream. The computational domain is depicted in figure \ref{fig:macrogeom}. Dirichlet boundary conditions on the velocity $(u_x,u_y)=(sin\alpha, cos\alpha)$ are imposed on the bottom and left sides of the domain, while a Neumann condition $\Sigma_{ij}n_j=0$ is imposed at the top and right sides of the domain. The no-slip condition $u_i=0$ is imposed on the surface of the inclusions $\partial\mathbb{M}$. For the solute concentration $c$, the Dirichlet condition $c=1$ applies on the left side of the domain, while $c=0$ is imposed on $\partial \mathbb{M}$. No-flux conditions apply on the external sides of the domain. The domain decomposition method \citep{Quarteroni2017} is employed to solve the macroscopic configuration by splitting the domain into two regions (cf. figure \ref{fig:macrogeom}b) connected by the membrane's homogenized condition \eqref{eq:macromodel} on $\mathbb{C}$. Continuity of velocity and stresses are imposed on the fluid-fluid interfaces, identified by the dashed lines in figure \ref{fig:macrogeom}b.  
The macroscopic simulations are solved iteratively using a fixed-point scheme: 
\begin{itemize}
    \item the macroscopic problem is solved using the Stokes solution;
    \item the values of $\check{u}_i,\tilde{u}_i$ (for the constant advection closure approach) or $\check{\Sigma}_{ij}^{\mathbb{U,D}},\tilde{\Sigma}_{ij}^{\mathbb{U,D}}$ (for the variable advection closure approach) are sampled along the surface $\mathbb{C}$ cell-by-cell;
    \item we substitute these values into the microscopic problems \eqref{eq:solvconddirac};
    \item we solve again the macroscopic simulation to get an updated version of the closure advection;
    \item we iterate the procedure until convergence.
\end{itemize}
The convergence criterion is satisfied when the difference between two subsequent computations in terms of $(u_i,c)$ on $\mathbb{C}$ is below $1\%$ of their mean values cell-by-cell. At each iteration, we perform $1/\epsilon$ microscopic (since the membrane is made of $1/\epsilon$ cells of length $l$, for a total length of $L$) and one macroscopic computation, with an averaging step in between. The comparison between the macroscopic model and the full-scale solution is performed in terms of averaged values of velocity and concentration on the membrane and point-wise values of the velocity, pressure and concentration fields far from the membrane. For simplicity, we compare separately the solvent and solute transport, which corresponds to considering $Re_L\neq 0, Pe_L=0$ (Section \ref{sub:momentum}) and $Re_L= 0, Pe_L\neq 0$ (Section \ref{sub:solute}), respectively. 
\subsection{Mass and momentum conservation} \label{sub:momentum}
\begin{figure}
    \centering
    \includegraphics[width=\textwidth]{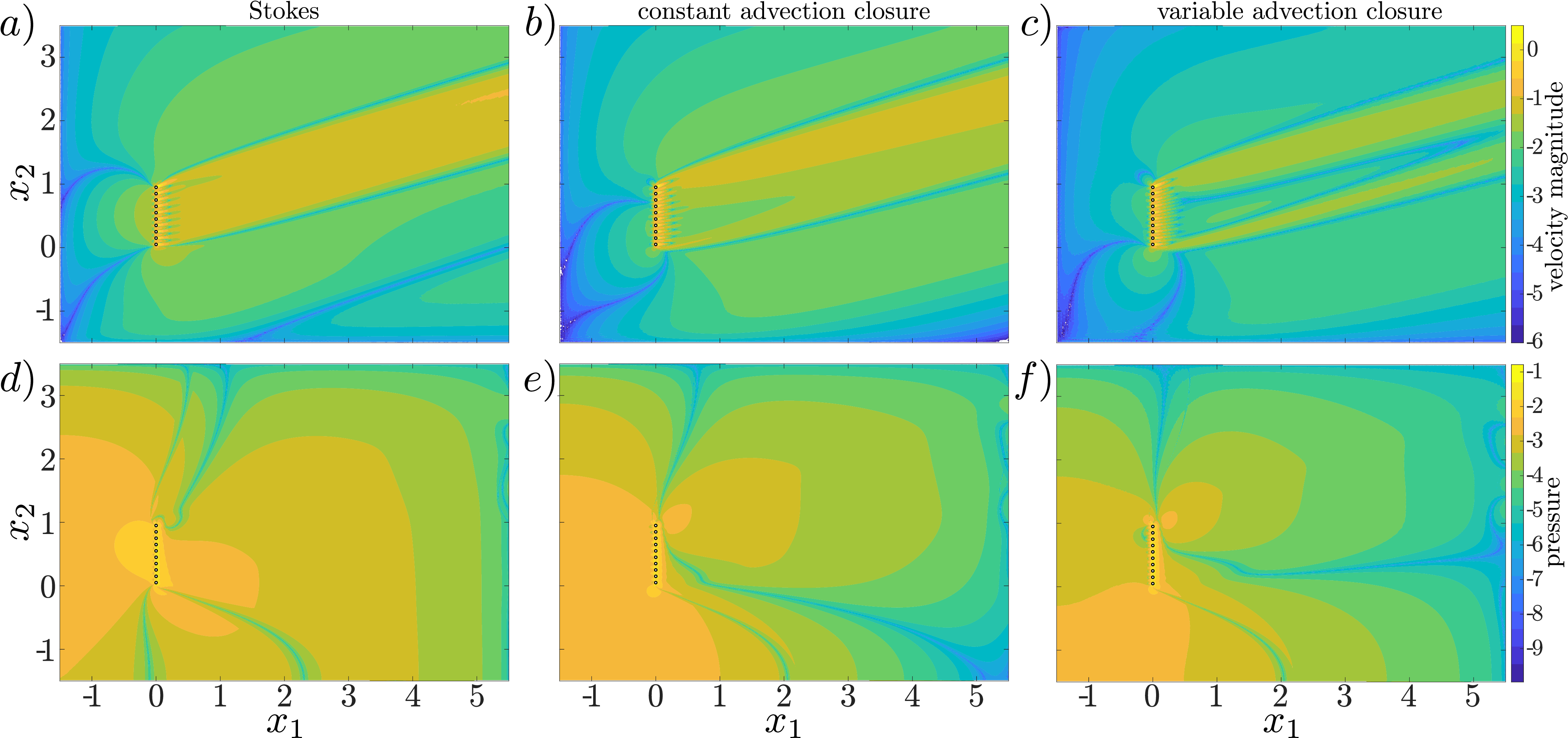}
    \caption{Contours of the relative error $e_r$ of ($a,b,c$) velocity magnitude and ($d,e,f$) pressure between the full-scale solution and the macroscopic models. Panels $(a,d)$ refer to Stokes models while $(b,e)$ and $(c,f)$ to the constant and variable advection closure, respectively. Fluid-flow and geometry parameters are $\alpha=75^{\circ}$, $\epsilon=0.1$ and $Re_L=400$. Colorbar is in $log_{10}e_r(\cdot)$-scale.}
    \label{fig:errorfull-scale solutionvs_momentum}
\end{figure}

\begin{figure}
    \centering
    \includegraphics[width=\textwidth]{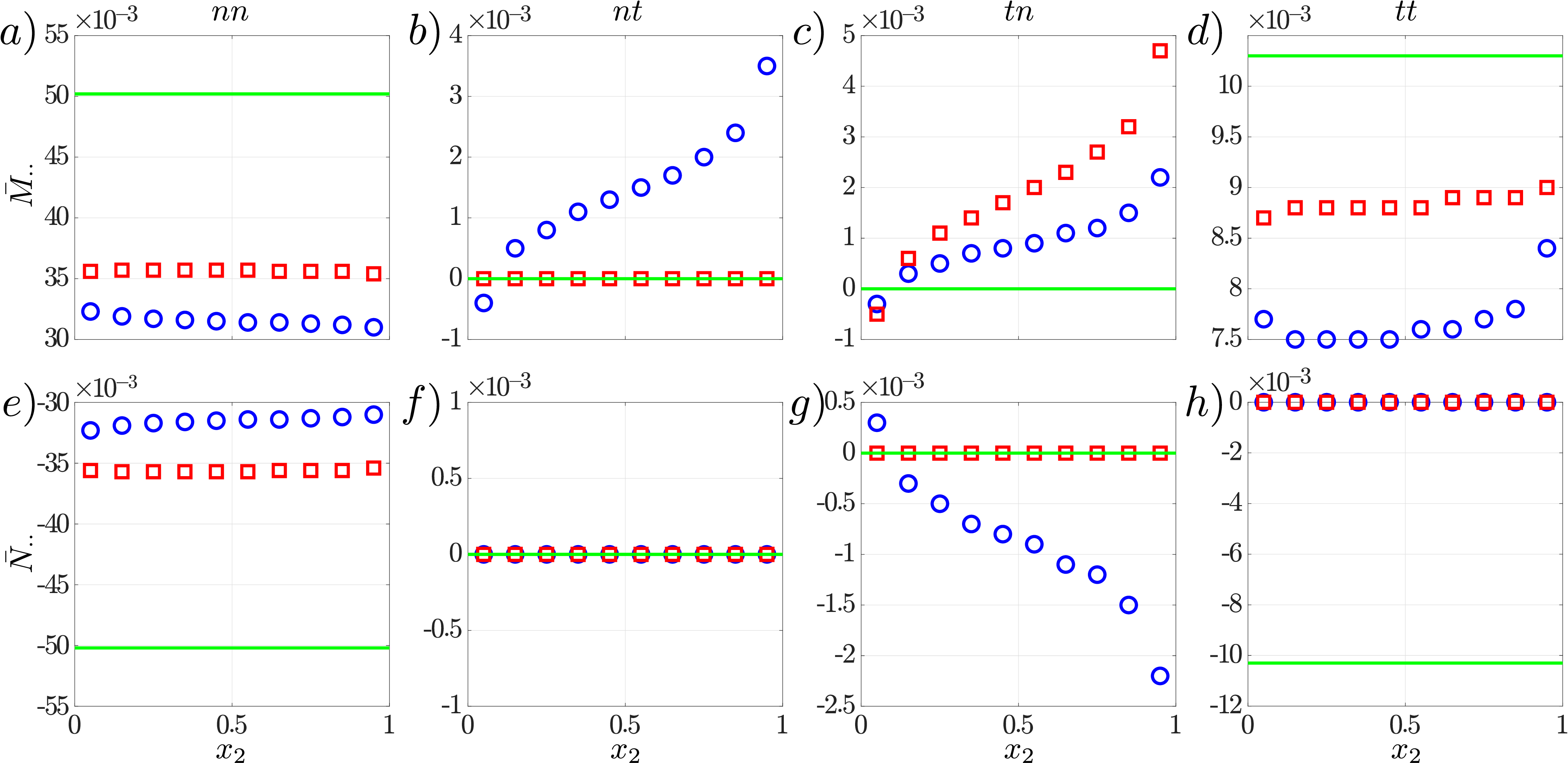}
    \caption{Averaged tensor values $\bar{M}_{nn}$ ($a$), $\bar{M}_{nt}$ ($b$), $\bar{M}_{tn}$ ($c$), $\bar{M}_{tt}$ ($d$), $\bar{N}_{nn}$ ($e$), $\bar{N}_{nt}$ ($f$), $\bar{N}_{tn}$ ($g$) and $\bar{N}_{tt}$ ($h$) for $Re_L=400$, $\alpha=75^{\circ}$, $\epsilon=0.1$ with the Stokes (green line), constant advection closure (blue circles) and variable advection closure (red squares) problems.}
    \label{tab:table MNRe400}
\end{figure}

\begin{figure}
    \centering
    \includegraphics[width=\textwidth]{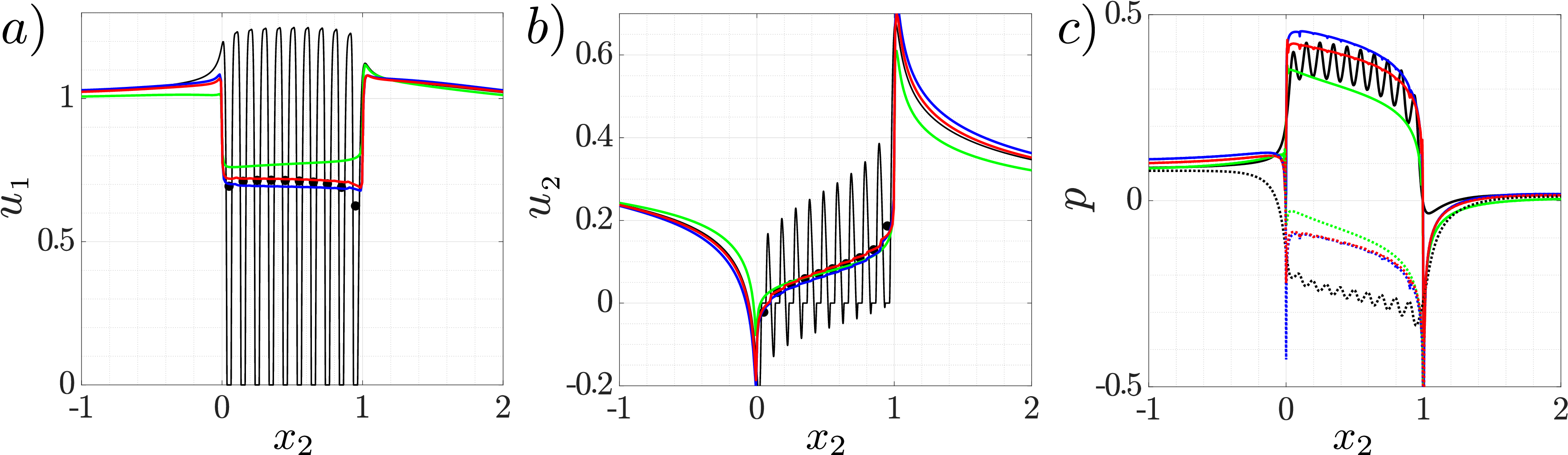}
    \caption{Horizontal ($a$) and vertical ($b$) velocity and pressure ($c$). In panels $a,b$, velocity values are sampled on the membrane centerline ($x_1=0$, $x_2\in[-1,2]$) in the full-scale solution and on the fictitious interface in the macroscopic simulations. In panel $c$ the pressure values are sampled at $x_1=\pm\epsilon/2$ and presented in dotted and full lines, respectively. Colour code: full-scale solution (black lines), averaged full-scale solution (black dots), constant (blue) and variable advection closure (red). Configuration parameters are $\alpha=75^{\circ}$, $\epsilon=0.1$ and $Re_L=400$.}
    \label{fig:uv_membrane_momentum}
\end{figure}

\begin{figure}
    \centering
    \includegraphics[width=\textwidth]{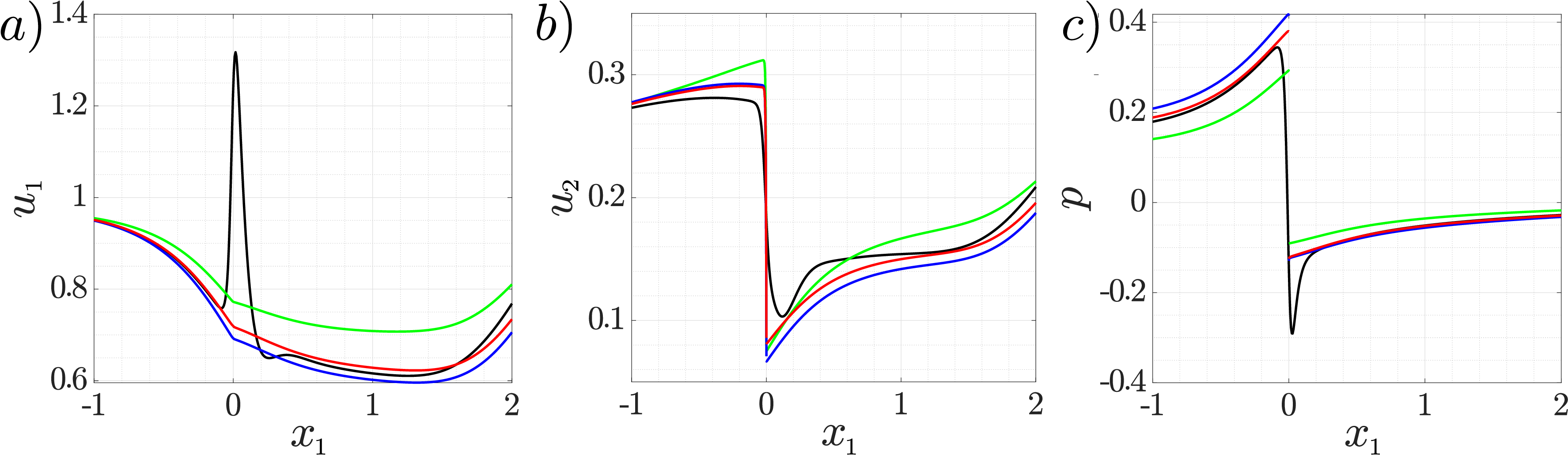}
    \caption{Horizontal ($a$) and vertical ($b$) velocity and pressure ($c$) sampled along a line at $x_2=0.5$ and $x_1\in[-1,2]$. The same colour code and configuration parameters as in figure \ref{fig:uv_membrane_momentum} are used.}
    \label{fig:uv_axis_momentum}
\end{figure}

In this section, we focus on the solvent flow with $Re_L=400$, $\epsilon=0.1$ and $\alpha=75^{\circ}$.
\begin{figure}
    \centering
    \includegraphics[width=\textwidth]{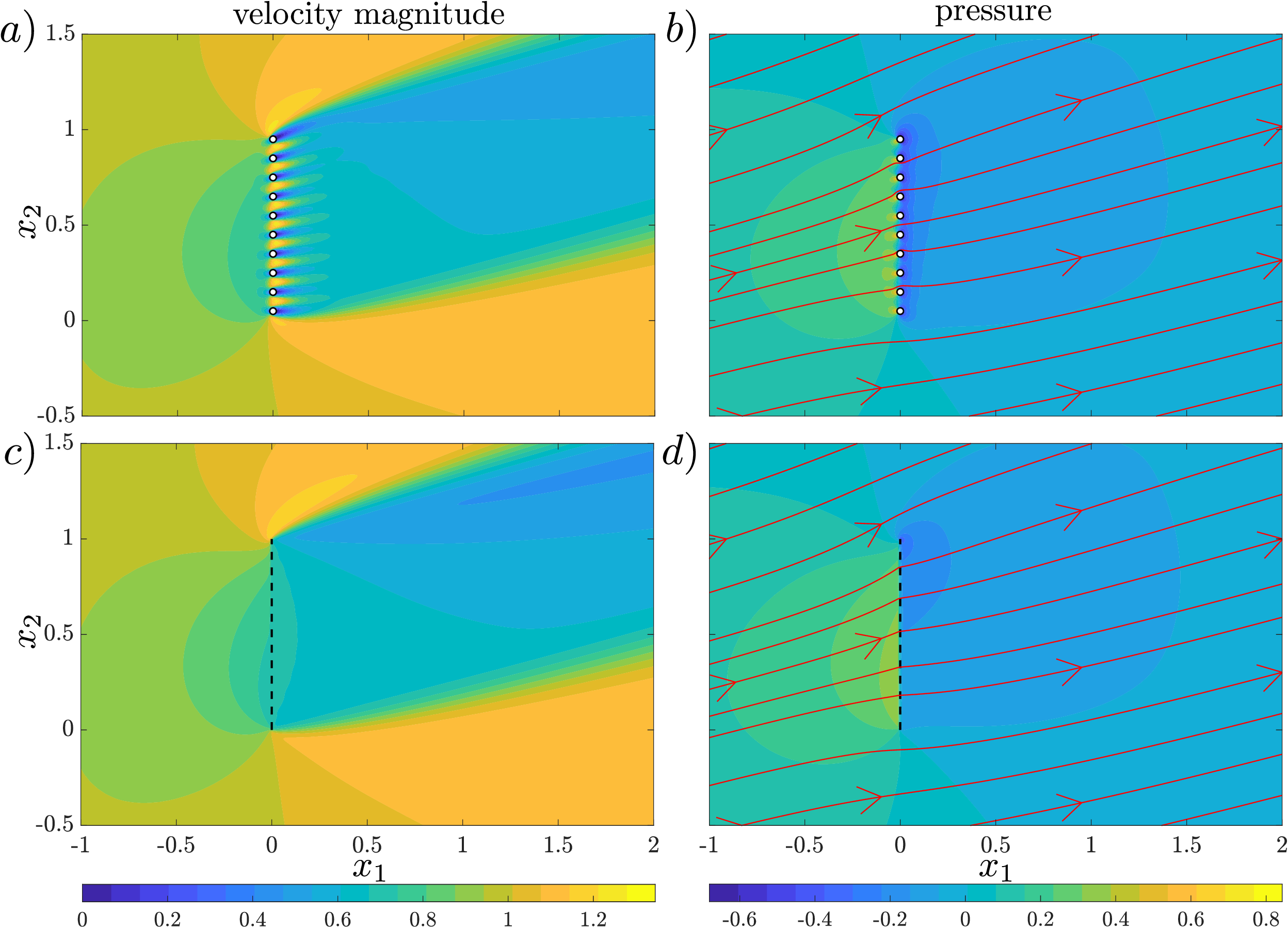}
    \caption{Velocity magnitude (panel $a,c$) and pressure iso-contours with velocity streamlines superimposed ($b,d$) in the proximity of the membrane for a full-scale ($a,b$) and a variable-advection macroscopic ($c,d$) simulation for $\alpha=75^{\circ}$, $\epsilon=0.1$ and $Re_L=400$. Black dashes indicate the location of the fictitious interface.}
    \label{fig:full-scale solution_momentum}
\end{figure}
In figure \ref{fig:full-scale solution_momentum}, we observe a wake developing downstream of each inclusion, forming a macroscopic wake downstream of the membrane, with parabolic-like velocity profiles across the openings of the membrane. Figure \ref{fig:errorfull-scale solutionvs_momentum} shows the relative differences in the fields between the full-scale and homogenized models. The largest discrepancies are found in the region immediately downstream of the membrane. 
These discrepancies decrease from the Stokes to the constant and variable advection closure models (frames $a,c$ and $b$, respectively), confirming that the quasi-linear models well capture the flow structures for non-negligible inertial effects and the variable advection model is a faithful approximation of the Navier-Stokes solution at the microscopic scale. Figure \ref{tab:table MNRe400} shows the values of $\bar{M}_{ij}$ and $\bar{N}_{ij}$ sampled on the membrane. The difference between the values of $\bar{M}_{ij}$ found using the Stokes and the constant and variable advection models is evident, in particular for the off-diagonal components. When advection is considered, the components $\cdot_{nn}$ and $\cdot_{tt}$ show a nearly constant value along the membrane which is different from the value predicted by the Stokes model.
A comparison of the velocity fields at the membrane centerline $\mathbb{C}$ is presented in figure \ref{fig:uv_membrane_momentum} and confirms the accuracy observed in figure \ref{fig:errorfull-scale solutionvs_momentum}. 
We consider also two sampling lines at $x_1=\pm \epsilon/2$, i.e. on the two sides of the interface, presented in figure \ref{fig:uv_membrane_momentum}c. 
Pressure, sampled at $x_1=\pm\epsilon/2$, is captured more accurately by the constant advection and the variable advection models than by the Stokes model. Figure \ref{fig:uv_axis_momentum} shows local comparisons on the $x_2=0.5$ lines, exhibiting a good agreement between the full-scale and macroscopic fields far from the membrane.

\begin{figure}
    \centering
    \includegraphics[width=\textwidth]{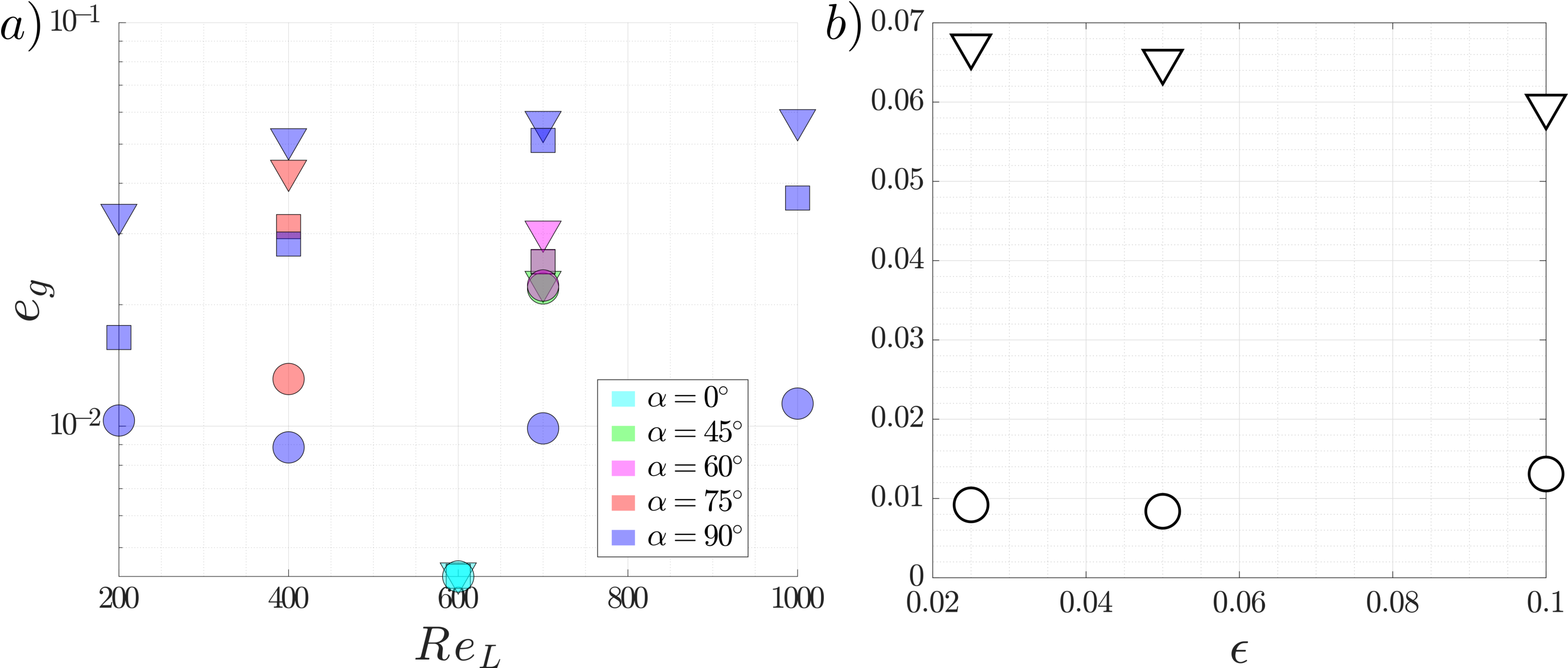}
    \caption{Panel $a$: global error $e_g$ defined in equation \eqref{eq:eg} for macroscopic simulations having $\epsilon=0.1$ and $\alpha$ ranging from $\alpha=0^{\circ}$ to $90^{\circ}$ and $Re_L$ from $200$ to $1000$. The values of $\alpha$ are represented using the colour code (legend in panel $a$), while the marker indicates the type of macroscopic model used to compute that data point (triangles for Stokes, squares for constant advection closure and circles for variable advection closure). Panel $b$: $e_g$ for the Stokes and variable advection closure model for different values of $\epsilon$. All simulations have  $Re_l=75$ and $\alpha=90^{\circ}$. A constant value of $Re_l=75$ has been realized by choosing the $(\epsilon,Re_L)$ couples as $(0.1,750)$, $(0.05,1500)$ and $(0.025,3000)$. Same marker code as for panel $a$.}
    \label{fig:general_errorplots}
\end{figure}

To assess the robustness of the previous observations, we vary $Re_L$, $\epsilon$ and $\alpha$ and quantitatively evaluate the agreement between the full-scale solution and the macroscopic models via a global error defined as
\begin{equation}
    e_g=\sqrt{\bar{e}_r(||\bm{u}||)^2+\bar{e}_r(|p|)^2}
\label{eq:eg}\end{equation}
where $\bar{e}_r$ is the mean of the point-wise relative error $e_r$ between the considered macroscopic model and the full-scale solution in the computational domain. Figure \ref{fig:general_errorplots}$a$ shows the errors calculated for several configurations such that $\epsilon=0.1$, $\alpha\in[0^{\circ},90^{\circ}]$ and $Re_L\in[200,1000]$. Going from Stokes to constant and variable advection models, an overall decrease of $e_g$ is noticed. Figure \ref{fig:general_errorplots}b presents the dependence of $e_g$ on the velocity at the membrane as a function of $\epsilon$ for $Re_l\approx \epsilon Re_L=75$. $\alpha=90^{\circ}$ for all computations. The error computed for the Stokes and the variable advection model shows nearly an order of magnitude of difference, with opposite trends as a function of $\epsilon$ (keeping $ \epsilon Re_L$ constant).
\subsection{Solute flux conservation} \label{sub:solute}
\begin{figure}
    \centering
    \includegraphics[width=\textwidth]{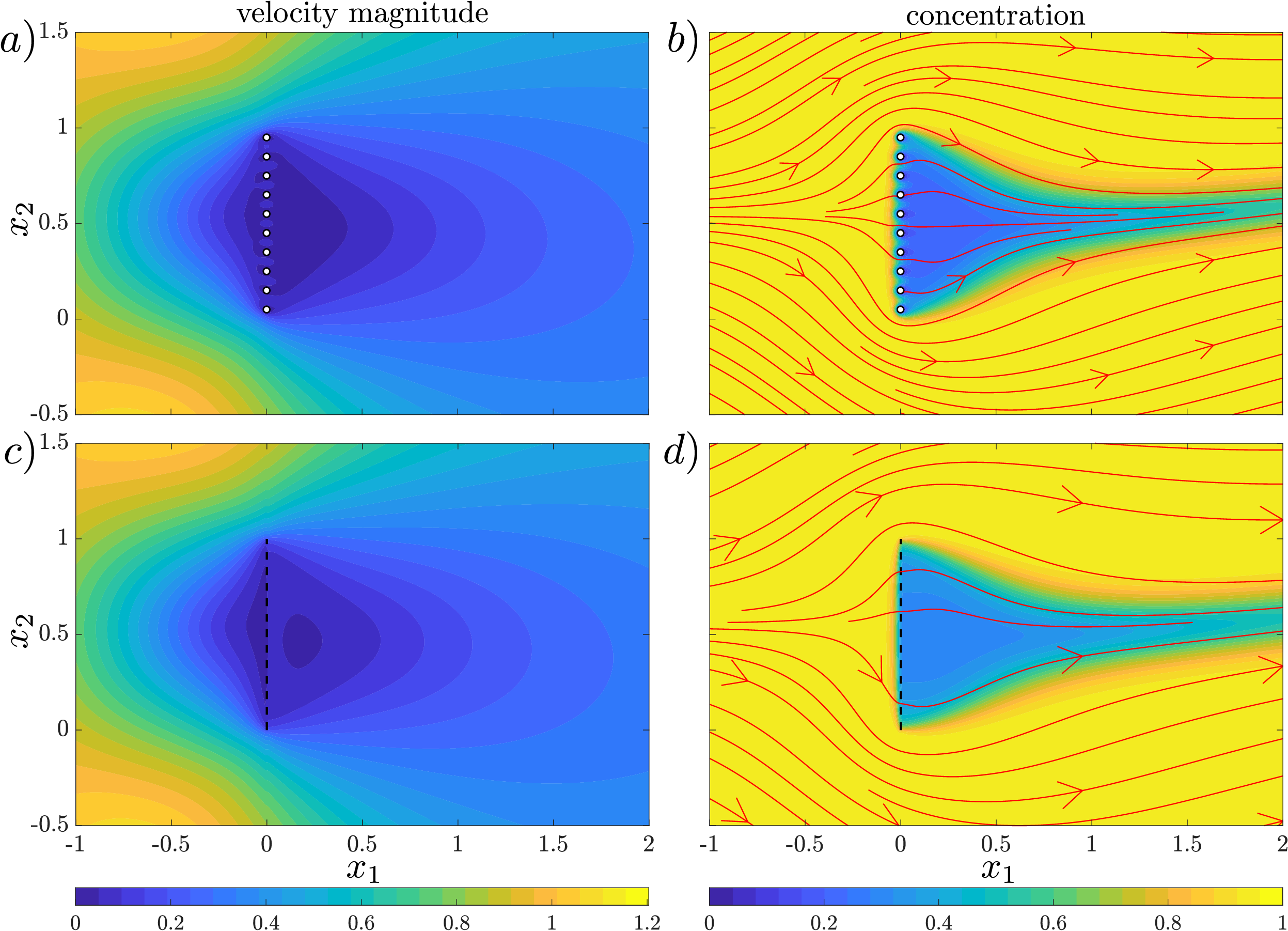}
    \caption{Velocity magnitude (panels $a,c$) and concentration iso-contours with velocity streamlines ($b,d$, red lines) for a full-scale ($a,b$) and a variable-advection macroscopic ($c,d$) solution at $Re_L=0$, $Pe_L=1000$ and $\alpha=90^{\circ}$ and $\epsilon=0.1$. Black dashes indicate the location of the fictitious interface.}
    \label{fig:full-scale solution_solute}
\end{figure}

\begin{figure}
    \centering
    \includegraphics[width=\textwidth]{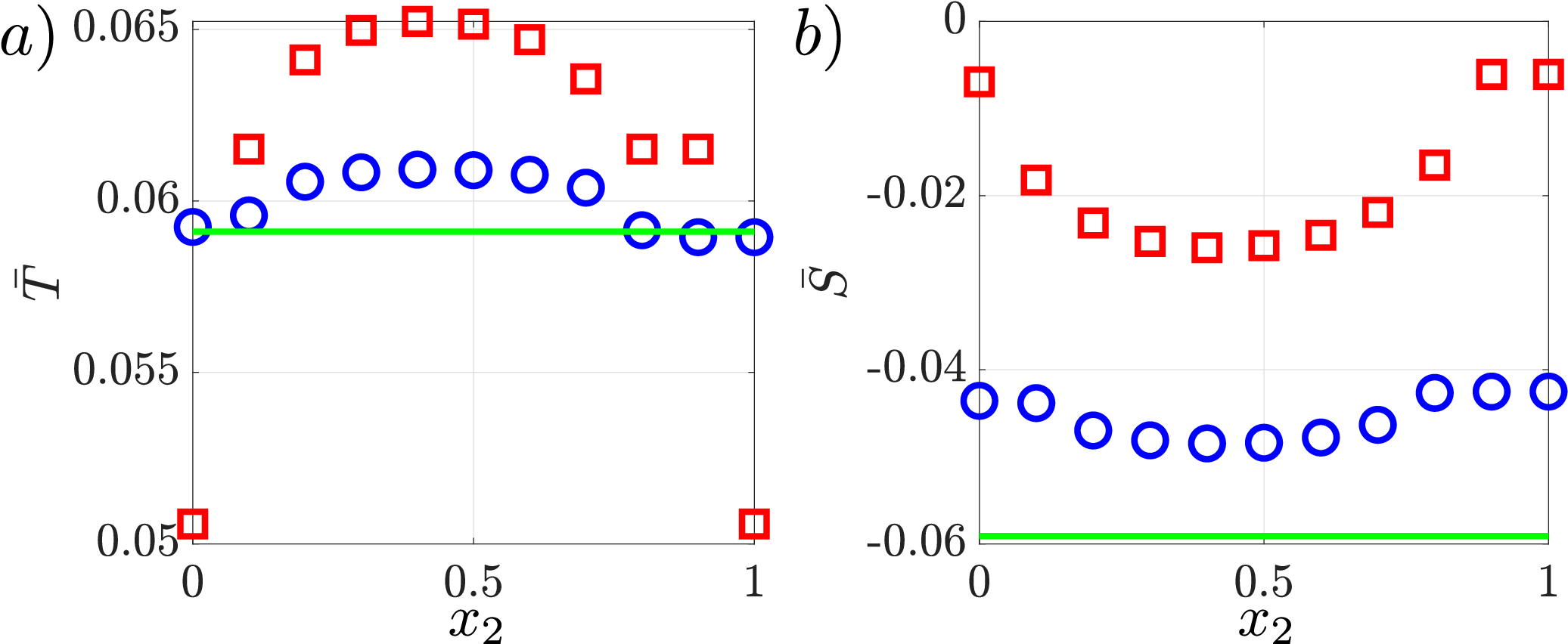}
    \caption{Values of $\bar{T}$ ($a$) and $\bar{S}$ ($b$) along the membrane for Stokes (green line), constant advection closure (blue circles) and variable advection closure (red squares). The flow parameters are $\alpha=90^{\circ}$, $\epsilon=0.1$, $Pe_L=1000$ and $Re_L=0$.}
    \label{fig:tableTS}
\end{figure}

We consider the case of $Pe_L> 0$. To increase $Pe_L$ we decrease the diffusivity $D$ while $Re_L$ remains negligible. This allows us to assess the reliability of the model independently of the solvent flow approximation. We consider a setup with $\alpha=90^{\circ}$, $\epsilon=0.1$ and $Pe_L=1000$. The iso-contours of the flow fields solved at all scales are presented in figure \ref{fig:full-scale solution_solute}. The velocity field (panel $a$) shows a small wake downstream of the membrane. The macroscopic flow does not present a re-circulation region, coherently with the hypothesis of negligible inertia. Conversely, the effect of the finite Péclet number is evident in panel $b$, where the solute concentration iso-levels are aligned with the solvent flow streamlines.
The $\bar{T}$ and $\bar{S}$ values found in the present flow configuration using the different models are collected in figure \ref{fig:tableTS}. The values of $\bar{T}$ and $\bar{S}$ found using the Stokes and constant advection model are quite similar, as opposed to the variable advection model.

\begin{figure}
    \centering
    \includegraphics[width=\textwidth]{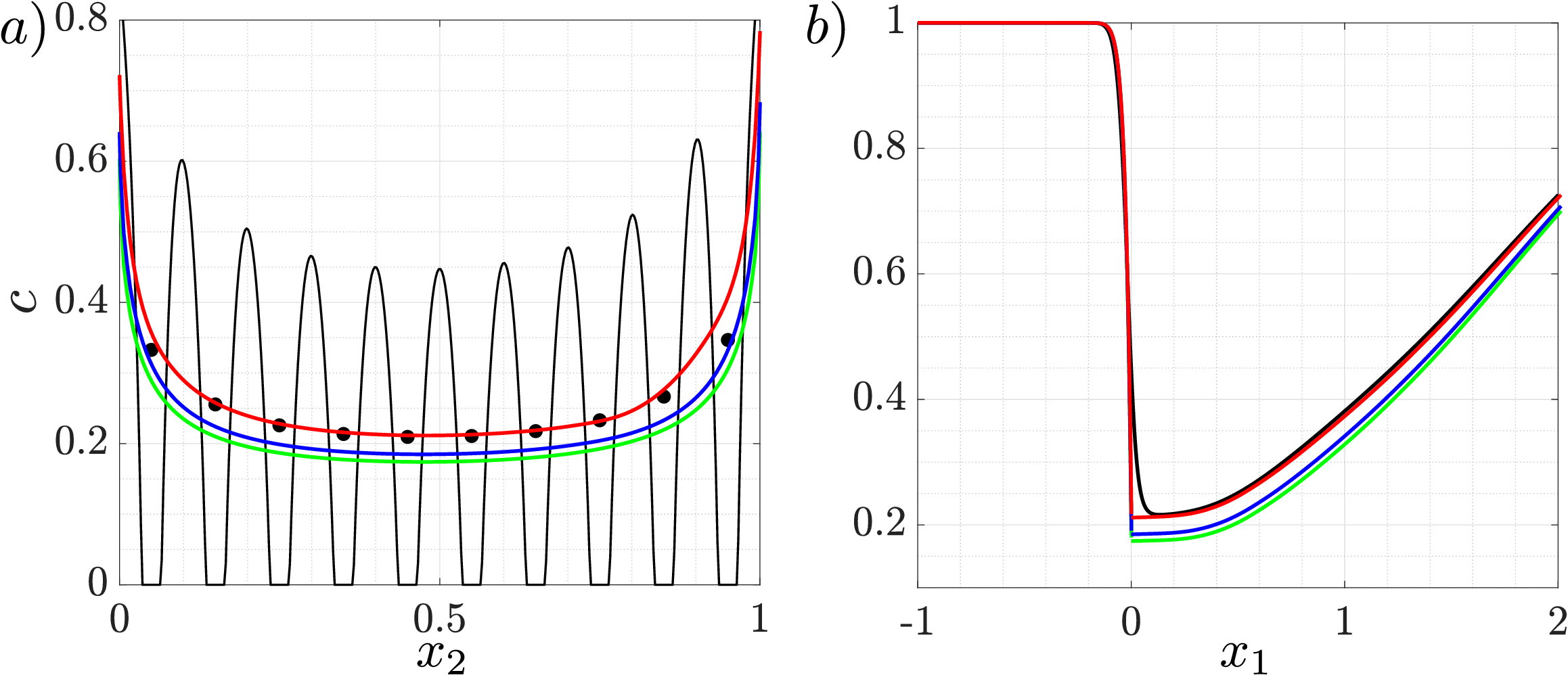}
    \caption{Comparison between the values of concentration $c$ sampled on the membrane centerline $\mathbb{C}$ ($a$) or on the membrane axis, corresponding to a line at $x_2=0.5$ and $x_1\in [-1,2]$ ($b$). The same colour coding as in figure \ref{fig:uv_membrane_momentum} has been adopted. The flow parameters are $\alpha=90^{\circ}$, $\epsilon=0.1$, $Pe_L=1000$ and $Re_L=0$.}
    \label{fig:errfull-scale solutionvs_solute}
\end{figure}

The variable advection model shows larger differences with respect to the Stokes solution compared to the constant advection closure model. 
We present a comparison of the concentration values on the membrane ($x_1=0$) and its axis ($x_2=0.5$) in figure \ref{fig:errfull-scale solutionvs_solute}. The variable advection closure approach shows a good agreement with the full-scale solution both on the membrane (panel $a$) and in the far-field, represented by the values of solute concentration sampled on the membrane axis (panel $b$). The constant advection closure offers little improvement in terms of accuracy, compared to the Stokes case.
We conclude that the maximum accuracy for the case of finite Péclet number is found using the variable advection closure, in analogy with the non-zero Reynolds number case. The quasi-linear homogeneous model is thus more accurate than the Stokes model. In addition, the variable advection approach shows a better agreement with full-scale simulations than the constant advection approach. However, we observe that the new macroscopic approximations are computationally more expensive than the Stokes model. Models of engineering interest used in preliminary design phases need to provide a fast and accurate output which can be used as a starting point for more expensive computations. In the following section, we discuss the trade-off between accuracy and computational efficiency, a key aspect of industrial flow modelling.

%% file: 3C-computat_efficiency.tex
\section{Towards data-driven homogenization: improving the computational efficiency through a central-value approximation}
\label{sec:computeffic}
The constant and variable advection closure models require the solution of problem \eqref{eq:solvconddirac} in each microscopic cell forming a single membrane, i.e. $1/\epsilon$ times at each iteration (or $N/\epsilon$ for $N$ membranes whose inclusions have spacing $\epsilon$). This leads to a loss in computational efficiency compared to the Stokes model, which requires only one microscopic solution without any iterative procedure. In the following, we propose a strategy to reduce the computational cost of the macroscopic solution focusing on the previously validated variable-advection approach. To avoid overloading this article, we consider only the hydrodynamic problem. However, the efficient solution strategy proposed in this section applies also to the constant advection model and advection-diffusion equations.
\subsection{Computing a single membrane using its central values} \label{sub:singlecentral}
\begin{figure}
    \centering
    \includegraphics[width=\textwidth]{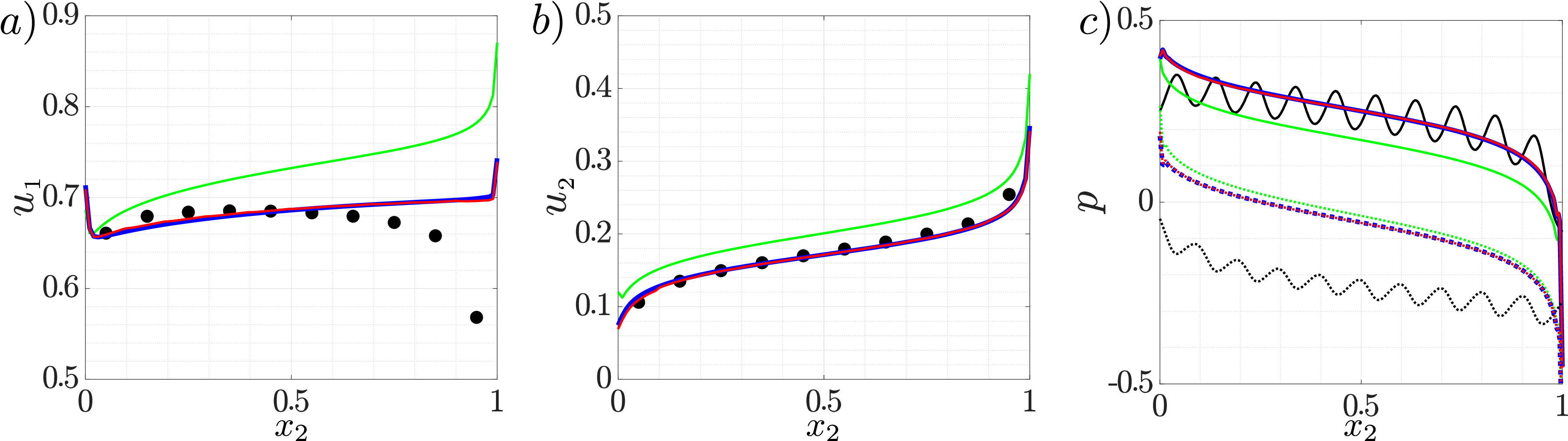}
    \caption{Comparison between the averaged full-scale solution normal velocity ($a$), tangential velocity ($b$) and pressure ($c$). Velocity components are sampled on the membrane centerline $\mathbb{C}$ in the full-scale solution and the fictitious interface in the macroscopic cases ($x_1=0, y\in[0,1]$). Pressure is sampled on two lines parallel to $\mathbb{C}$ and distant $\pm \epsilon/2$ (dotted and full lines respectively). The colour code is common to all panels: black dots for the cell-averaged values of the full-scale solution, green lines for the Stokes model, red lines for the variable advection closure unclustered model, and blue lines for the variable advection closure, clustered model. The flow and geometry parameters are $\alpha=60^{\circ}$, $Re_L=700$, $\epsilon=0.1$.}
    \label{fig:clustercellwise_mem}
\end{figure}

\begin{figure}
    \centering
    \includegraphics[width=\textwidth]{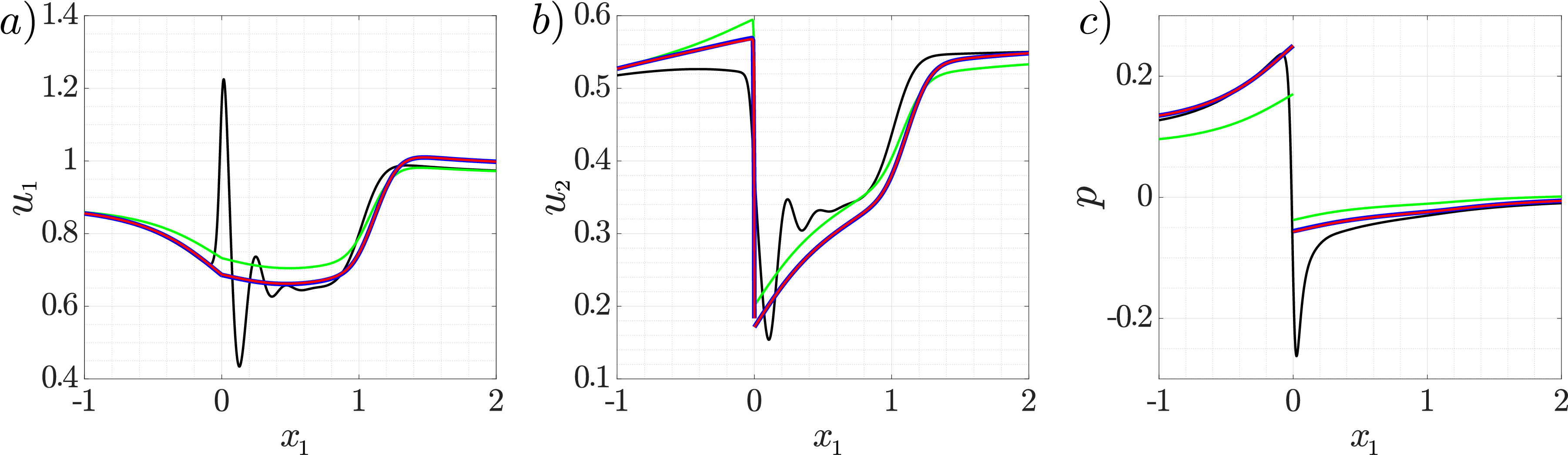}
    \caption{Comparison between flow field values of normal velocity ($a$), tangential velocity ($b$) and pressure ($c$) sampled along the axis of the membrane, $x_2=0.5, x_1\in[-1,2]$. The same colour code as in figure \ref{fig:clustercellwise_mem} is adopted. The flow and geometry parameters are $\alpha=60^{\circ}$, $Re_L=700$, $\epsilon=0.1$.}
    \label{fig:clustercellwise_axis}
\end{figure}

The quantities $\mathcal{P}=\left(\check{u}_i,\tilde{u}_i,\check{\Sigma}^{\mathbb{U},\mathbb{D}}_{ij},\tilde{\Sigma}^{\mathbb{U},\mathbb{D}}_{ij}\right)$ affect the values of $\bar{M}_{ij}$, $\bar{N}_{ij}$, $\bar{T}$ and $\bar{S}$. We notice that the dispersion of $\bar{M}_{ij}$ and $\bar{N}_{ij}$ observed in figure \ref{tab:table MNRe400} is small (less than $5\%$ of the mean membrane values for the three dominant components ($\bar{M}_{nn},\bar{N}_{nn}$ and $\bar{M}_{tt}$) of the $\bar{M}_{ij}$ and $ \bar{N}_{ij}$ tensors). For the simple geometry presented in figure \ref{fig:macrogeom}, the central value of this dispersion roughly corresponds to the values of $\mathcal{P}$ evaluated at the membrane's centre. We can thus employ the values of $\mathcal{P}$ reached in the middle of the membrane in the evaluation of $\bar{M}_{ij},\bar{N}_{ij},\bar{T}$ and $\bar{S}$. To assess the validity of this approach, we consider the case $\alpha=60^{\circ}, Re_L=700$. Figure \ref{fig:clustercellwise_mem} presents a comparison of the cell-averaged values of the solvent velocity and pressure on the membrane obtained in the full-scale solution and using the Stokes and variable advection closure models, clustered (i.e. obtained using a single microscopic computation for all cells at each iteration) and unclustered (i.e. using one microscopic computation for each cell at each iteration). 
Figure \ref{fig:clustercellwise_axis} compares the solvent flow fields sampled along the axis of the membrane. In both figures, the variable advection closure clustered and unclustered versions predict very similar values of the flow fields.  Average relative differences between the two approaches are of order $0.05\epsilon$, well below the accuracy of the homogeneous model ($\mathcal{O}(\epsilon)$). In terms of computational cost, the clustered version requires only one computation per iteration (similarly to the Stokes solution), while the un-clustered one requires $1/\epsilon$.
\subsection{Clustering the cells} \label{sub:clustering}
\begin{figure}
    \centering
    \includegraphics[width=\textwidth]{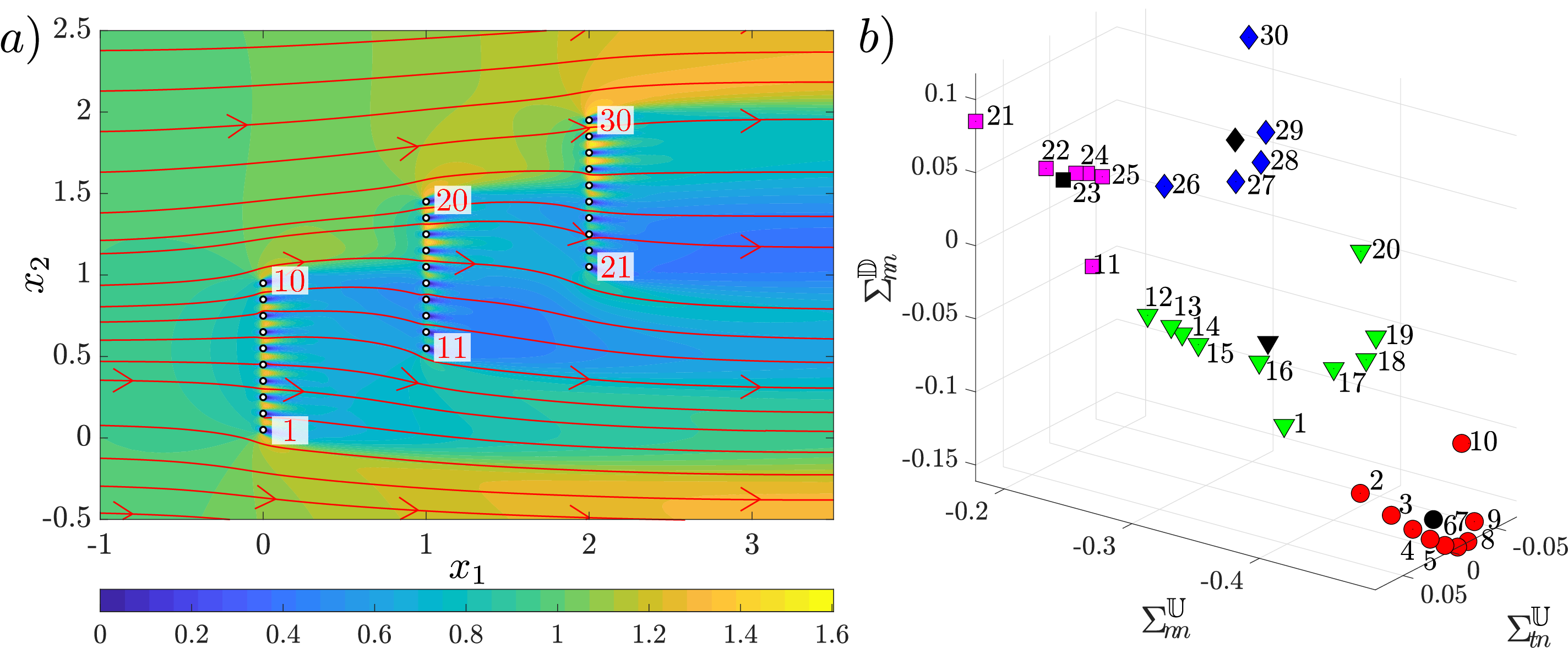}
    \caption{Panel $a$: contours of velocity magnitude and streamlines (red) in the full-scale solution for $\alpha=90^{\circ}$, $\epsilon=0.1$ and $Re_L=500$. Each of the three identical membranes contains ten circular solid inclusions of porosity $0.7$ and spacing $\epsilon$. Inclusions $1,11$ and $21$ are located in $(0,0),(L,0.5L)$ and $(2L,L)$, respectively. The domain extent is $x_1\in[-1.5L,5.5L]$ and $y\in[-1.5L,3.5L]$. No slip is imposed on the top and bottom sides of the domain, while $(u_1,u_2)=(1,0)$ is imposed on the leftmost side and $\Sigma_{ij}n;j=0$ on the rightmost side of the domain. Cells are numbered from bottom to top and left to right (red text). Panel $b$: a possible clustering choice for the cells on the membrane obtained using the Matlab k-Means algorithm. Each cluster is visualized in a different marker and colour. Black markers represent the centroids of each cluster. $\Sigma_{tn}^{\mathbb{D}}$ is close to zero, thus it does not affect the results and it is hence not shown. Black numbers refer to corresponding cells in panel $a$.}
    \label{fig:fieldscluster}
\end{figure}

\begin{figure}
    \centering
    \includegraphics[width=\textwidth]{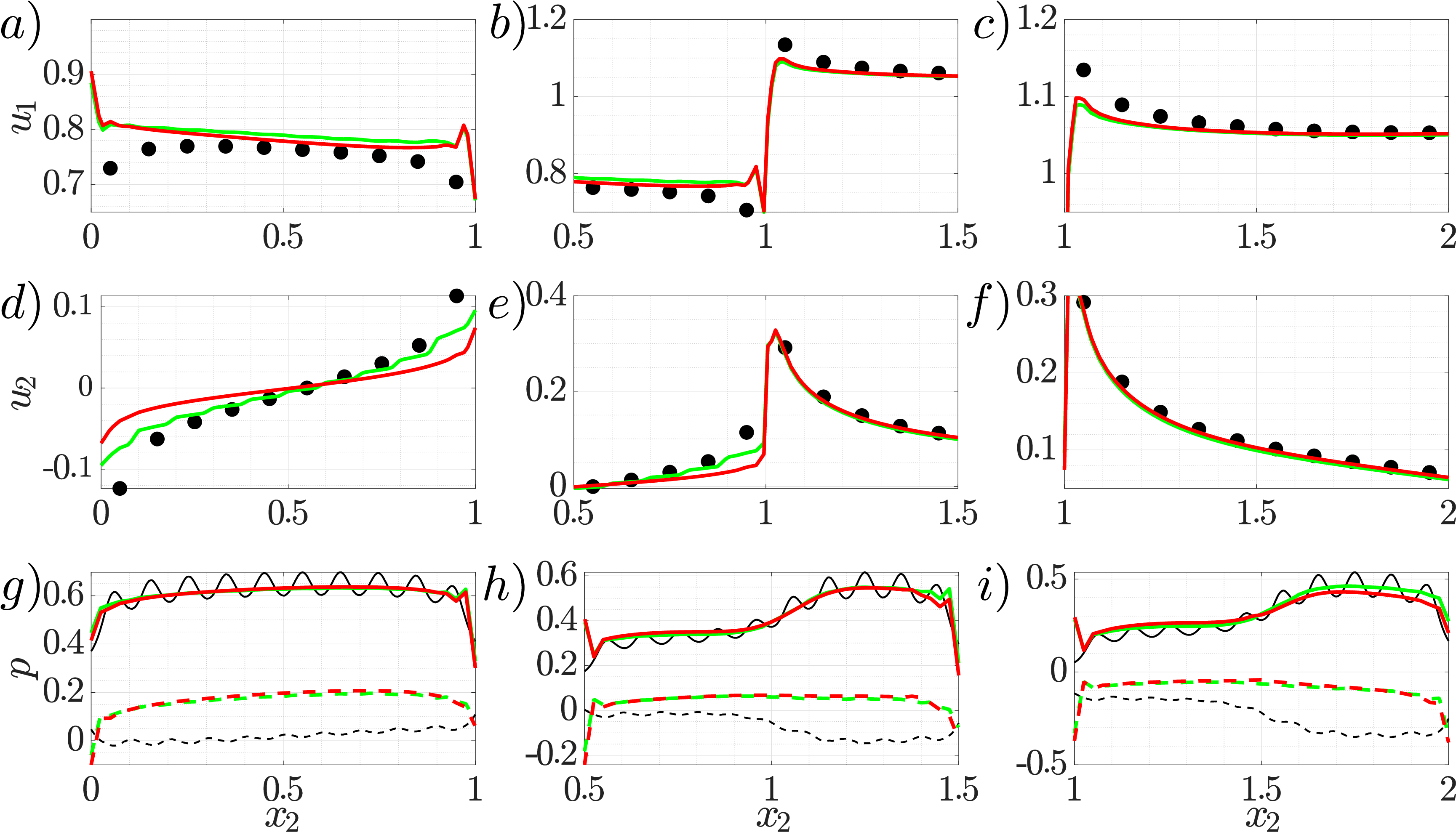}
    \caption{Normal ($a-c$) and tangential ($d-f$) velocity components on the centerline of each membrane. Black dots represent cell-averaged values of the full-scale solution, green lines represent the unclustered variable advection closure model and red lines are the clustered variable advection closure model. In panels $g-i$, we collect pressure values sampled at $\pm\epsilon/2$ from each centerline in the full-scale solution (full black line) and corresponding values for the unclustered variable advection closure model (green) and clustered variable advection closure (red) sampled on the corresponding sides of the fictitious interfaces. Panels $a,d,g$ refer to the leftmost membrane, $b,e,h$ to the central one and $c,f,i$ to the rightmost membrane in figure \ref{fig:fieldscluster}a.}
    \label{fig:clustermembrane}
\end{figure}

In the previous paragraph, the centre of the membrane was used to compute one set of approximate values of the tensors $\bar{M}_{ij}$ and $\bar{N}_{ij}$ in place of $1/\epsilon$ (i.e. one for each microscopic cell). Flow and geometry configurations of industrial interest are generally more complicated. We introduce a clustering algorithm in our macroscopic simulation workflow to automate the choice of a subset of flow conditions (i.e. of the $\mathcal{P}$ quantities) which can approximate the real, cell-wise distribution of $\mathcal{P}$. Given a set of length $N$ of the flow quantities $\mathcal{P}$, we divide $\mathcal{P}$ in $K \leq N$ clusters in which each element belongs to the cluster with the nearest centroid. We thus represent the cluster using its centroid. Several algorithms for splitting datasets into clusters have been proposed in the literature, see \cite{BishopML} for a review. Testing their relative performance in the present case goes beyond the scope of this work. For this reason, as an example, we choose the K-Means++ \citep{kmeans}, implemented in Matlab 2023a. This algorithm splits a given set of data into a user-defined number of clusters according to a notion of distance, Euclidean in the present case. The optimal number of clusters can be found using some heuristic procedures, like the ``elbow rule'' \citep{tibshirani}, but the cluster distribution is sufficiently evident in this case that the iterative procedure can be avoided. Figure \ref{fig:fieldscluster}a shows the considered flow configuration. 
It consists of three identical membranes composed of circular inclusions with spacing $\epsilon=0.1$, immersed in a free stream at $Re_L=500$. Velocity magnitude iso-levels and velocity streamlines are presented in panel $a$. On the top and bottom sides of the domain, a no-slip boundary condition is applied. The leftmost membrane is fully exposed to the flow, as well as the top portions of the other two membranes, while the bottom portions experience milder flow conditions. Panel $b$ shows that the data is clustered in three or four blocks, corresponding to the different flow conditions at the cell level. By computing the microscopic cases using the $\Sigma_{ij}^{\mathbb{U,D}}$ of each cluster centroid, we solve the corresponding macroscopic computation. A comparison of the velocity at the membranes in the full-scale solution and the clustered variable advection closure macroscopic solution is presented in figure \ref{fig:clustermembrane}. The prediction based on the four clusters is satisfactory compared to the full-scale solution, with minor discrepancies only in the tangential velocity of the red cluster.
In terms of required computational resources, on a laptop with one INTEL i9-10900HK CPU (8 cores, 2.40GHz) and 32GB of RAM, a microscopic variable advection closure simulation has a run-time of up to 3 minutes while a macroscopic simulation takes about 1 minute. The clustering step proposed in this section has allowed for a reduction in the number of microscopic problems from $30$ to $4$ cases per iteration, which translates in a microscopic iteration step $7.5$ times faster. In general, if we suppose that the number of iterations and the number of clusters at each iteration are $\mathcal{O}(1)$, while the number of microscopic cases to run at each iteration without clustering is $\mathcal{O}(1/\epsilon)$, then clustering makes the iterative procedure $\mathcal{O}(1/\epsilon)$ times faster. For the cases proposed in this work, a few iterations were needed for convergence: for $\epsilon Re_L$ of order $10$, generally, $1-2$ iterations are sufficient, while for $\epsilon Re_L$ of order $100$, we need $5-6$ iterations.  

%% file: 4-conclusion.tex
\section{Conclusion and perspectives}\label{sec:conclu}
In this work, we developed a model to predict the solvent flow and diluted solute transport across thin permeable membranes in the case of non-negligible inertia at the pore scale. We exploited the separation of scales between the membrane and the pore size to decompose the mathematical problem into purely macroscopic equations and a microscopic problem which requires parametric inputs from the macro-scale stemming from the flow inertia at the pore scale. The presence of inertia therefore requires the approximation of the advective terms in the equations for the solvent flow and solute transport. An iterative procedure is employed to feed the microscopic problems with the macroscopic quantities included in the inertial terms within the Navier-Stokes and advection-diffusion equations. The macroscopic solution converges in an average sense to the fully resolved direct numerical simulations on the membrane and in the far-field. 

Our work aims at extending the macroscopic, homogenized, description of filtration flows across permeable membranes toward practical applications when the Reynolds and Peclet numbers at the pore scale cannot be neglected. A relevant improvement in accuracy compared to the inertialess version is found, for microscopic Reynolds number of order $10$. In addition, preliminary faithful results can be obtained through machine learning clustering algorithms. This approach could offer a beneficial trade-off between efficiency and accuracy for large systems of industrial interest, such as filters or fuel cells \citep{Cullen2021, Wang2023}. In typical applications involving microfluidic circuits, flow inertia is either disregarded or considered a detrimental effect since it may affect the low-Reynolds number hydrodynamic analytical predictions. However, our model establishes a specific relation between permeability properties and flow inertia. Hence, filtration properties can be finely tuned by selecting the flow rate within microfluidic channels, using the Reynolds and Péclet numbers as control parameters, and thus becoming a design parameter in addition to geometrical properties. As a matter of fact, for the same geometry, one can obtain a broad spectrum of filtration properties by simply changing the flow rate. Flow inertia thus becomes an opportunity to extend the working conditions of filtering systems.

This work could be extended in several ways. First, in the present form, the model cannot be applied to the case of macroscopic unsteady configuration triggered by pore-scale hydrodynamic instabilities \citep{nicolle_eames_2011}. To extend the model toward larger Reynolds numbers, space-time averages need to be introduced and data-aided physics-informed microscopic models are required to efficiently handle the micro-macro coupling. The quasi-linear iterative strategy developed in the present paper may open the door to the modelling of other, low-Re, non-linear phenomena, such as osmotic or phoretic flows through semi-permeable structures, which typically present a strong non-linear coupling between the solvent velocity and the solute concentration \citep{marbachbocquet2017}.

%% file: 5-acknowledgements.tex
\section{Acknowledgements}
This work was supported by the Swiss National Science Foundation (grant no. $PZ00P2\_193180$). The authors are profoundly grateful to Prof. F. Gallaire for his assistance and, in particular, for the insightful discussions concerning the treatment of the non-linearities in the microscopic problem.
\section{Conflict of interest}
The authors declare no conflict of interest.

%% file: 6-code.tex
\section{Code availability}
The source code, along with some examples, is available on GitHub at \href{https://github.com/kevlfmi/QuasiLinearHomogenization.git}{this link.}

%% file: Appendix.tex
\appendix
\section{Computational details} \label{app:computat_details}
\begin{figure}
    \centering
    \includegraphics[width=0.9\textwidth]{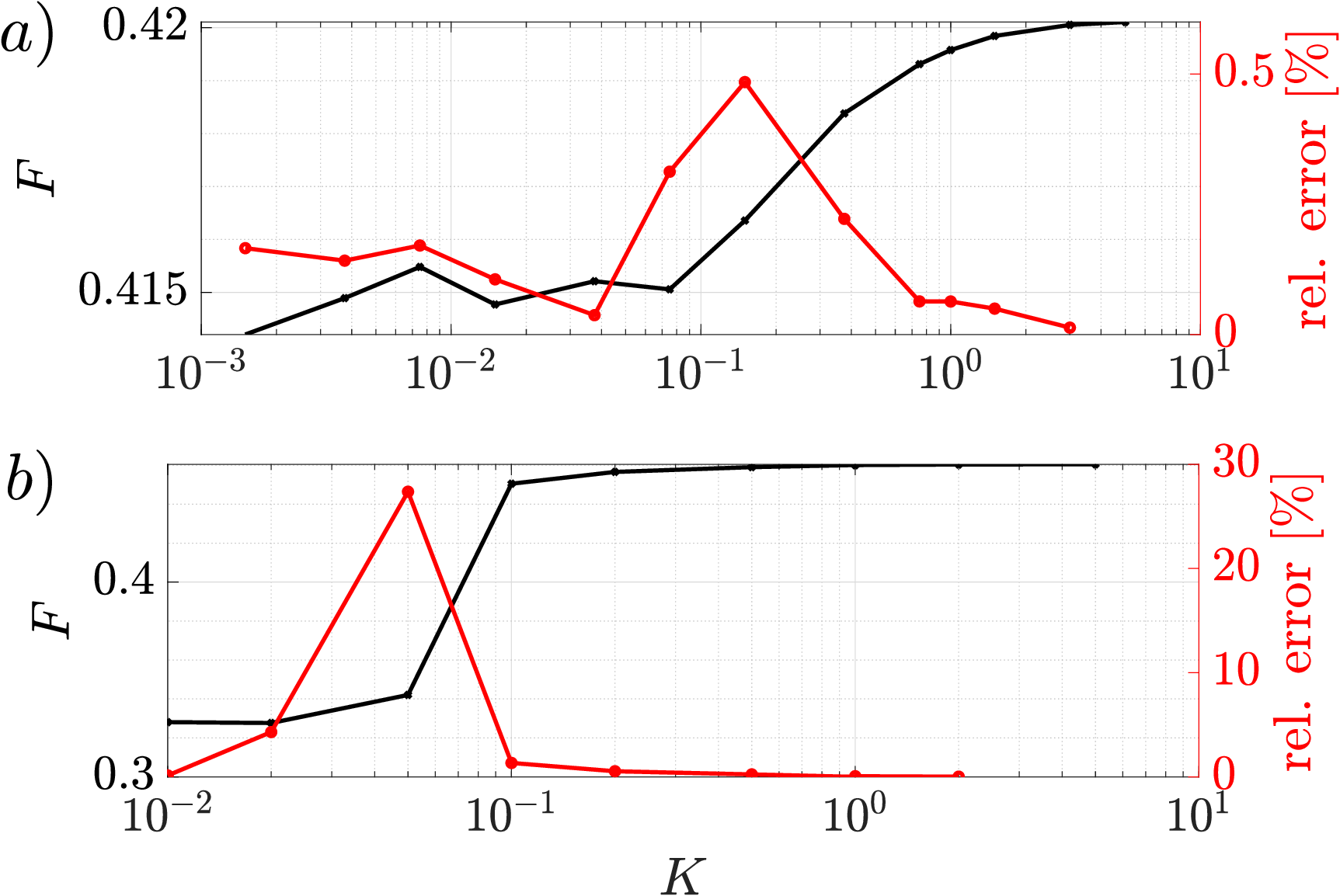}
    \caption{Force magnitude $F$ (black) acting on $\mathbb{C}$ in the macroscopic simulation (panel $a$) and on the solid boundary $\partial\mathbb{M}$ in the full-scale simulation (panel $b$) and relative error between two consequent values of $F$ (red) as a function of $K$ for $Re_L=500,\alpha=90^{\circ}, \epsilon=0.1$ and circular solid inclusions of porosity $0.7$.}
    \label{fig:meshconvergence4_2}
\end{figure}

The microscopic and macroscopic equations and the full-scale solution presented in \S \ref{sec:fullscalemacro} have been solved numerically using the finite-element solver COMSOL Multiphysics 6.0. The coupling between the microscopic and the macroscopic problems is automated thanks to the MatLab Livelink extension. The spatial convergence of the mesh has been tested for each model and flow configuration. The same criterion has been used for the microscopic calculations. All simulations are performed using a Taylor-Hood P2-P1 scheme for coupling velocity and pressure and a P2 scheme for the solute concentration. We present hereafter the results of the convergence study for $\epsilon=0.1, Re_L=500, \alpha=90^{\circ}$ and circular inclusions with porosity $0.7$ for the macroscopic and full-scale simulations. We start with a mesh whose typical size is 
\begin{itemize}
    \item $0.01L$ on the macroscopic interface and $0.045L$ far from it for the macroscopic problem (corresponding to $K=1$ in figure \ref{fig:meshconvergence4_2}a),
    \item $0.0015L$ on the solid boundary and $0.05L$ far from it for the full-scale problem (corresponding to $K=1$ in figure \ref{fig:meshconvergence4_2}b),
    \item $0.04l$ near the solid inclusion and $0.5l$ far from them for the microscopic problems (corresponding to $K=1$ in figure \ref{tab:meshconvergence3}),
\end{itemize}
and explore a range of $K$ values to verify that the results are mesh-independent, where $K$ is a parameter dividing the initial mesh sizing for each simulation. We consider the mesh converged when the force magnitude $F=||\int_{\mathbb{C}}\Sigma_{ij}n_j dS||$ applied by the fluid on the fictitious interface $\mathbb{C}$ in the macroscopic simulation and on $\partial\mathbb{M}$ in the full-scale simulation reaches an asymptotic value and the relative error between two $F$ values sampled at two subsequent values of $K$ is less than $0.2\%$. The converged macroscopic mesh has $91304$ elements and the full-scale one has $237900$, both corresponding to $K=1$ (cf. figure \ref{fig:meshconvergence4_2}). The parameters of the microscopic case are $\check{\Sigma}^{\mathbb{U}}_{ij}n_j=10^4$ and $\check{\Sigma}^{\mathbb{D}}_{ij}n_j=0$. We consider the mesh converged when the relative difference between the non-zero values of $\bar{M}_{ij}, \bar{N}_{ij}$ is less than $1\%$ for two subsequent values of $K$. The final mesh contains $12780$ elements, corresponding to $K=10$ in table \ref{tab:meshconvergence3}.
\begin{table}
    \centering
    \begin{tabular}{ccccccccc}
        $K$ & $\bar{M}_{nn}$ & $\bar{M}_{nt}$ & $\bar{M}_{tn}$ & $\bar{M}_{tt}$ & $\bar{N}_{nn}$ & $\bar{N}_{nt}$ & $\bar{N}_{tn}$ & $\bar{N}_{tt}$ \\ 
        100 & 1.453E-02 & 4.007E-05 & 5.045E-15 & -4.655E-15 & -1.453E-02 & -4.007E-05 & -2.208E-06 & -3.027E-03 \\ 
        50 & 1.453E-02 & 4.001E-05 & 2.402E-13 & -4.656E-14 & -1.453E-02 & -4.001E-05 & -2.322E-06 & -3.030E-03 \\ 
        10 & 1.454E-02 & 5.223E-05 & -6.720E-04 & -4.719E-05 & -1.454E-02 & -5.223E-05 & 2.318E-04 & -3.012E-03 \\ 
        5 & 1.454E-02 & 4.994E-05 & -5.528E-04 & -3.025E-05 & -1.454E-02 & -4.994E-05 & 1.902E-04 & -3.017E-03 \\ 
        2 & 1.454E-02 & 5.008E-05 & -5.637E-04 & -2.807E-05 & -1.454E-02 & -5.008E-05 & 1.939E-04 & -3.018E-03 \\ 
        1 & 1.294E-02 & 1.156E-03 & -4.071E-02 & -4.199E-03 & -1.294E-02 & -1.156E-03 & 1.196E-02 & -1.900E-03 \\ 
    \end{tabular}
    \vspace{0.5cm}
    \caption{Averaged values of $M_{ij}$ as a function of the mesh factor $K$ for a microscopic problem with variable advection computed around a circular solid inclusion of porosity $0.7$ for $\check{\Sigma}^{\mathbb{U}}_{ij}=10^4$ and $\check{\Sigma}^{\mathbb{D}}_{ij}=0$. Only the values of $\bar{M}_{nn}, \bar{N}_{nn}$ and $\bar{N}_{tt}$ are considered for the mesh convergence study since the other components have negligible values at convergence ($K\geq 10$).}
    \label{tab:meshconvergence3}
\end{table}
\vspace{0.5cm}

\section{Further variable advection closure maps} \label{app:maps}

\begin{figure}
    \centering
    \includegraphics[width=\textwidth]{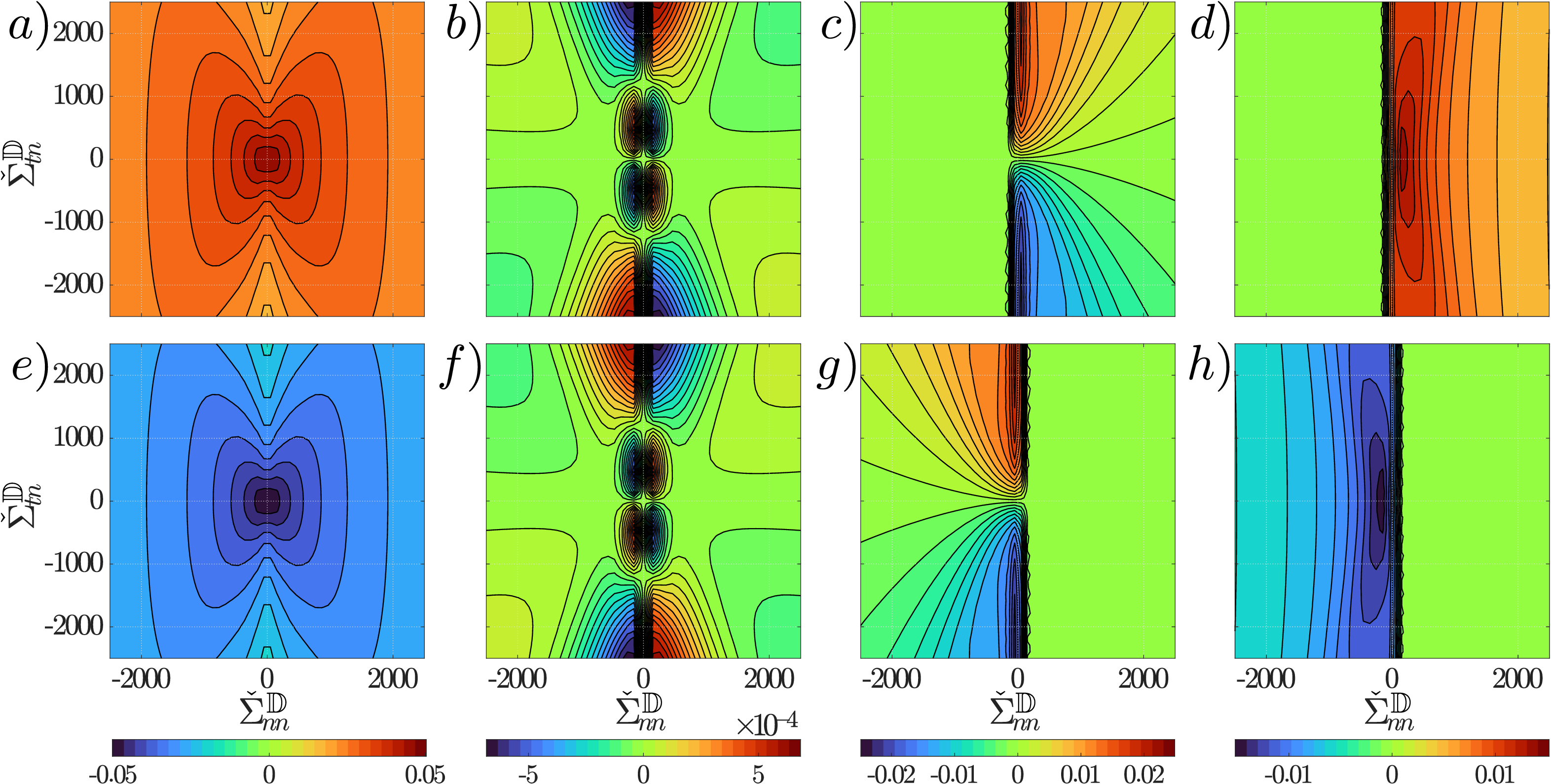}
    \caption{Average tensor values $\bar{M}_{nn}$ ($a$), $\bar{M}_{nt}$ ($b$), $\bar{M}_{tn}$ ($c$), $\bar{M}_{tt}$ ($d$), $\bar{N}_{nn}$ ($e$),  $\bar{N}_{nt}$ ($f$), $\bar{N}_{tn}$ ($g$), $\bar{N}_{tt}$ ($h$) varying $\check{\Sigma}^{\mathbb{D}}_{nn}$ and $\check{\Sigma}^{\mathbb{D}}_{tn}$, while $\check{\Sigma}^{\mathbb{U}}_{nn}=\check{\Sigma}^{\mathbb{U}}_{tn}=0$.}
    \label{fig:NSE_momentum_MAPS2}
\end{figure}

\begin{figure}
    \centering
    \includegraphics[width=\textwidth]{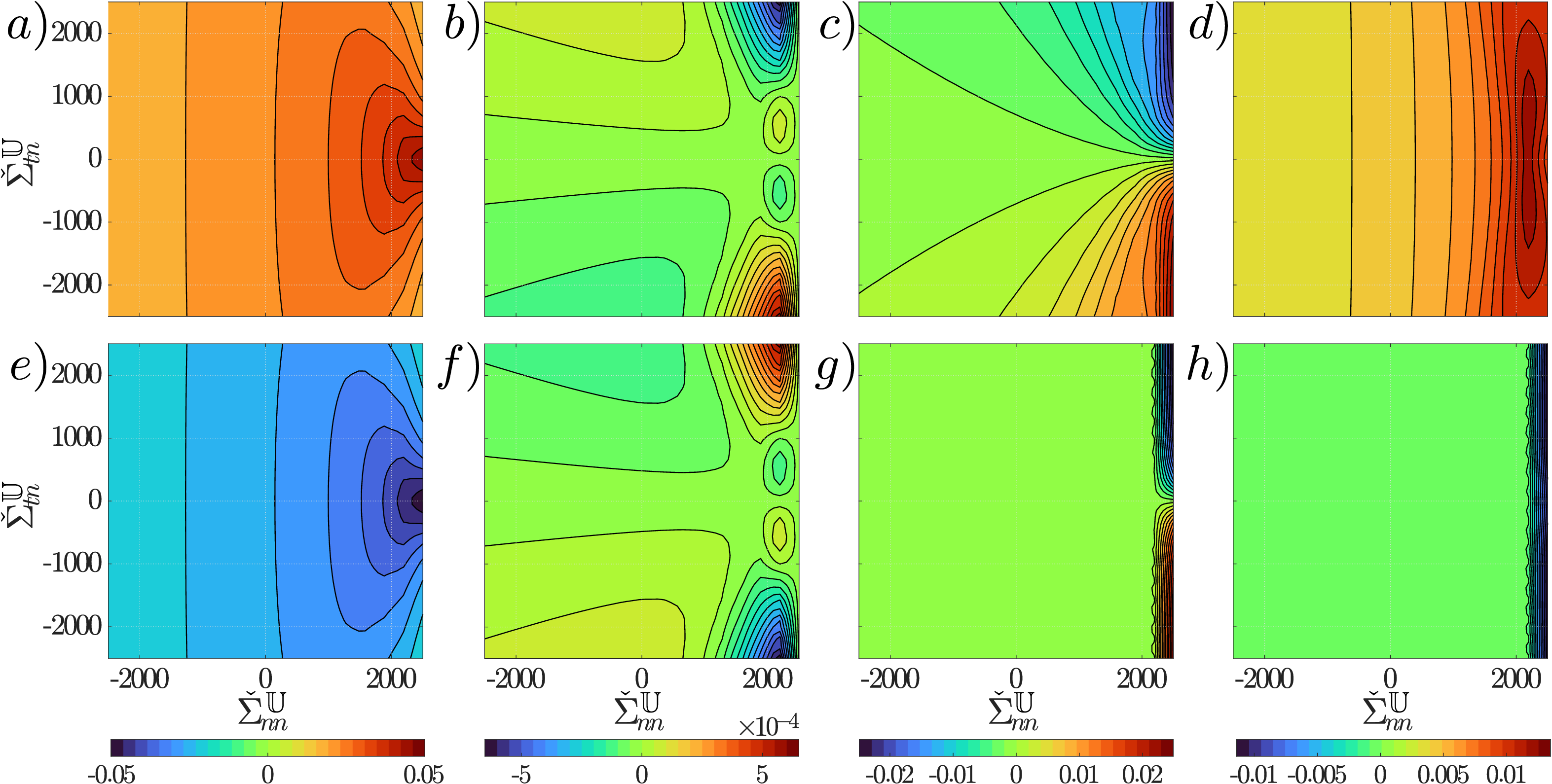}
    \caption{Average tensor values $\bar{M}_{nn}$ ($a$), $\bar{M}_{nt}$ ($b$), $\bar{M}_{tn}$ ($c$), $\bar{M}_{tt}$ ($d$), $\bar{N}_{nn}$ ($e$),  $\bar{N}_{nt}$ ($f$), $\bar{N}_{tn}$ ($g$), $\bar{N}_{tt}$ ($h$) varying $\check{\Sigma}^{\mathbb{U}}_{nn}=\epsilon^2 Re_L^2 \Sigma^{\mathbb{U}}_{nn}$ and $\check{\Sigma}^{\mathbb{U}}_{tn}$, while $\check{\Sigma}^{\mathbb{D}}_{nn}=2500$ and $\check{\Sigma}^{\mathbb{D}}_{tn}=0$.}
    \label{fig:NSE_momentum_MAPS_10}
\end{figure}

\begin{figure}
    \centering
    \includegraphics[width=\textwidth]{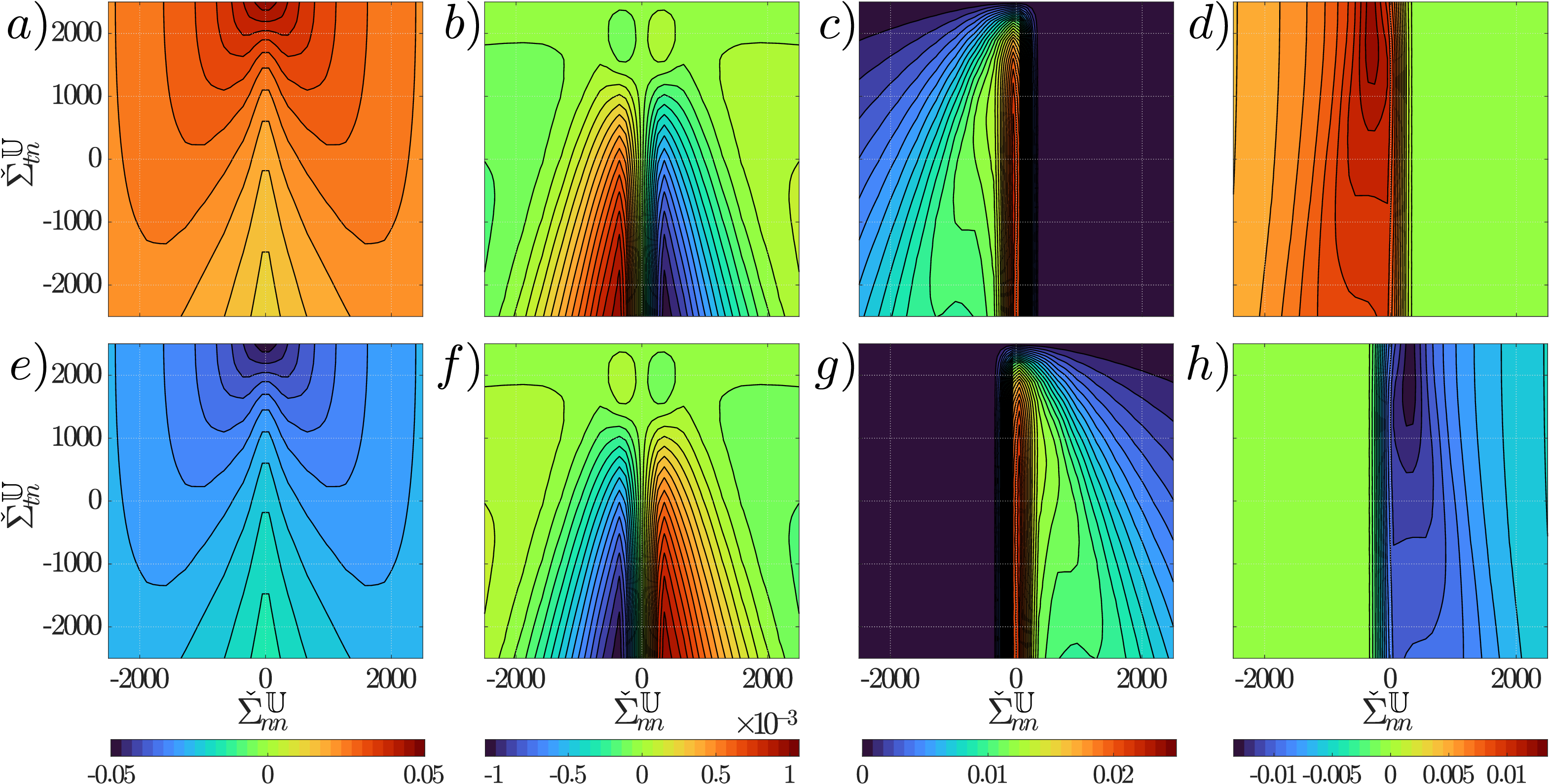}
    \caption{Average tensor values $\bar{M}_{nn}$ ($a$), $\bar{M}_{nt}$ ($b$), $\bar{M}_{tn}$ ($c$), $\bar{M}_{tt}$ ($d$), $\bar{N}_{nn}$ ($e$),  $\bar{N}_{nt}$ ($f$), $\bar{N}_{tn}$ ($g$), $\bar{N}_{tt}$ ($h$) varying $\check{\Sigma}^{\mathbb{U}}_{nn}=\epsilon^2 Re_L^2 \Sigma^{\mathbb{U}}_{nn}$ and $\check{\Sigma}^{\mathbb{U}}_{tn}$, while $\check{\Sigma}^{\mathbb{D}}_{nn}=0$ and $\check{\Sigma}^{\mathbb{D}}_{tn}=2500$.}
    \label{fig:NSE_momentum_MAPS_01}
\end{figure}

\begin{figure}
    \centering
    \includegraphics[width=\textwidth]{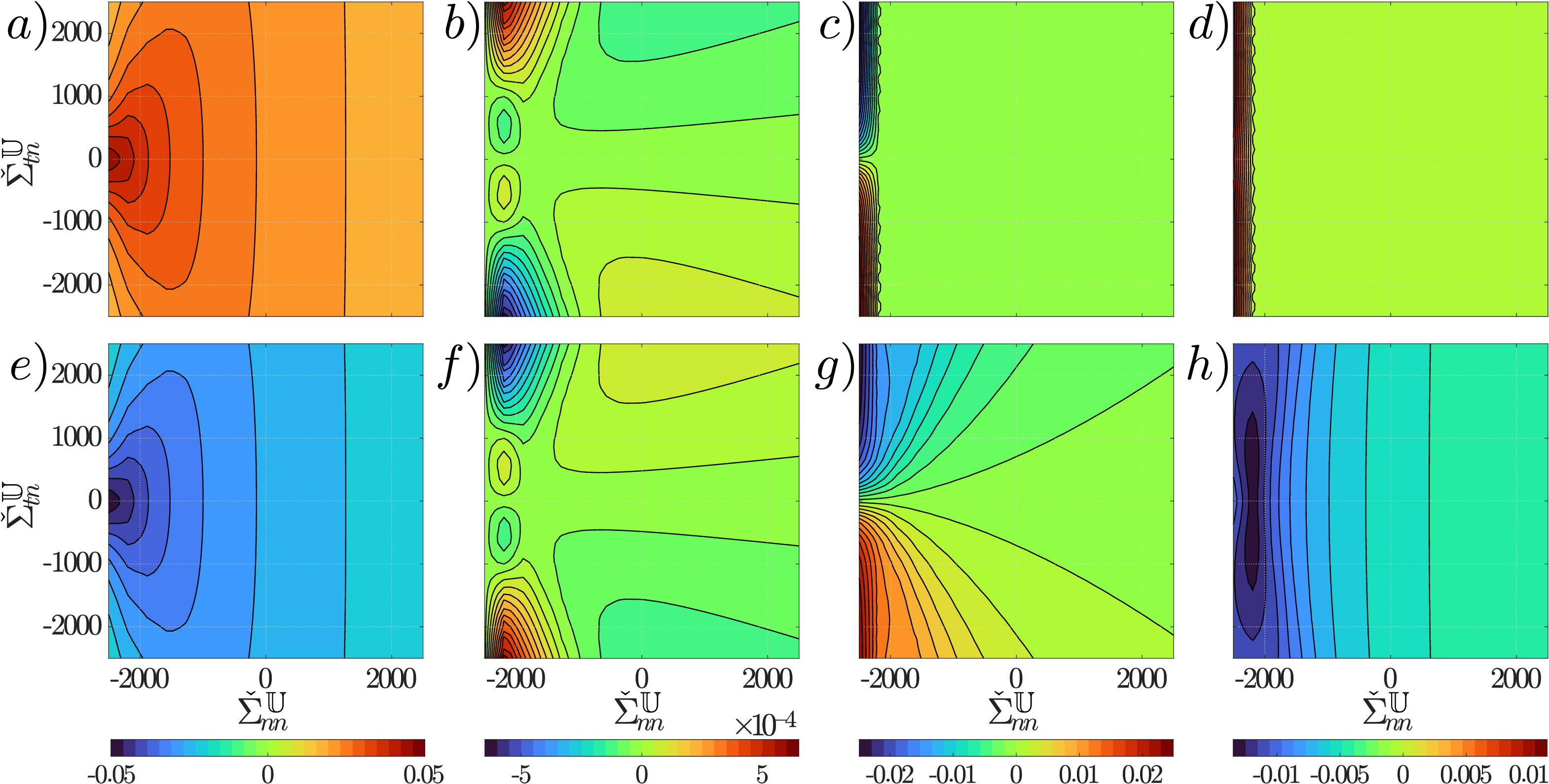}
    \caption{Average tensor values $\bar{M}_{nn}$ ($a$), $\bar{M}_{nt}$ ($b$), $\bar{M}_{tn}$ ($c$), $\bar{M}_{tt}$ ($d$), $\bar{N}_{nn}$ ($e$),  $\bar{N}_{nt}$ ($f$), $\bar{N}_{tn}$ ($g$), $\bar{N}_{tt}$ ($h$) varying $\check{\Sigma}^{\mathbb{U}}_{nn}=\epsilon^2 Re_L^2 \Sigma^{\mathbb{U}}_{nn}$ and $\check{\Sigma}^{\mathbb{U}}_{tn}$, while $\check{\Sigma}^{\mathbb{D}}_{nn}=-2500$ and $\check{\Sigma}^{\mathbb{D}}_{tn}=0$.}
    \label{fig:NSE_momentum_MAPS_-10}
\end{figure}

\begin{figure}
    \centering
    \includegraphics[width=\textwidth]{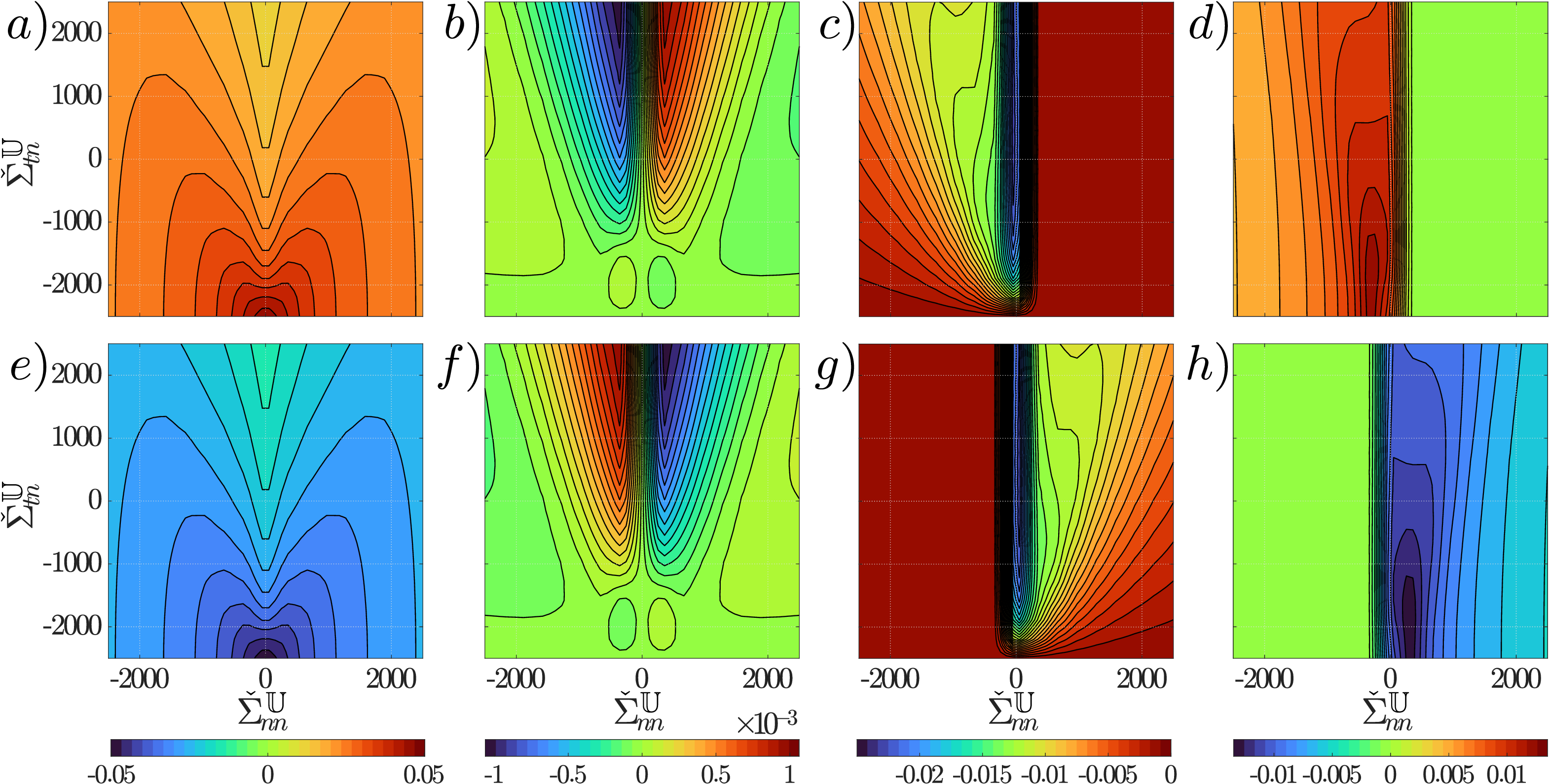}
    \caption{Average tensor values $\bar{M}_{nn}$ ($a$), $\bar{M}_{nt}$ ($b$), $\bar{M}_{tn}$ ($c$), $\bar{M}_{tt}$ ($d$), $\bar{N}_{nn}$ ($e$),  $\bar{N}_{nt}$ ($f$), $\bar{N}_{tn}$ ($g$), $\bar{N}_{tt}$ ($h$) varying $\check{\Sigma}^{\mathbb{U}}_{nn}=\epsilon^2 Re_L^2 \Sigma^{\mathbb{U}}_{nn}$ and $\check{\Sigma}^{\mathbb{U}}_{tn}$, while $\check{\Sigma}^{\mathbb{D}}_{nn}=0$ and $\check{\Sigma}^{\mathbb{D}}_{tn}=-2500$.}
    \label{fig:NSE_momentum_MAPS_0-1}
\end{figure}

Figures \ref{fig:NSE_momentum_MAPS2}-\ref{fig:NSE_momentum_MAPS_0-1} present further 2D sub-manifolds of the 4-dimensional manifold of $\bar{M}_{ij}$ and $\bar{N}_{ij}$ values found by solving the variable advection closure microscopic problem \eqref{eq:solvcondVAC}.
